\documentclass[10pt, conference, letterpaper]{IEEEtran}
\usepackage[letterpaper, left = 0.68in, right =0.68in, top = 0.7in, bottom = 1in]{geometry}

\def\arxivdisclaimer{1} 
\def\isshorter{0}
\def\review{0} 

\usepackage[noadjust]{cite}
\if\review1
\usepackage{todonotes}
\else
\usepackage[disable]{todonotes}
\fi
\usepackage[acronym]{glossaries}
\usepackage{amsthm}
\usepackage{amsmath, amssymb}
\usepackage{balance}

\usepackage{algorithm}
\usepackage{algpseudocode}

\usepackage{svg} 
\usepackage{graphicx} 
\usepackage{pgfplots} 
\usepackage{pgfplotstable} 
\usetikzlibrary{spy} 
\usetikzlibrary{arrows.meta} 
\usetikzlibrary{external}
\usepackage{xfrac} 
\usepackage{siunitx} 
\pgfplotsset{compat=newest} 
\pgfplotsset{
    /pgfplots/layers/Bowpark/.define layer set={
        axis background,axis grid,main,axis ticks,axis lines,axis tick labels,
        axis descriptions,axis foreground
    }{/pgfplots/layers/standard},
    colormap={jet}{
        rgb255(0cm)=(0,0,128);
        rgb255(1cm)=(0,0,255);
        rgb255(3cm)=(0,255,255);
        rgb255(5cm)=(255,255,0);
        rgb255(7cm)=(255,0,0);
        rgb255(8cm)=(128,0,0)
    }
} 

\pgfplotsset{
    colormap={jet_inue}{
        rgb=(1, 1, 1)
        rgb=(0.99804, 0.99804, 0.99905)
        rgb=(0.99609, 0.99609, 0.99813)
        rgb=(0.99413, 0.99413, 0.99725)
        rgb=(0.99217, 0.99217, 0.99639)
        rgb=(0.99022, 0.99022, 0.99557)
        rgb=(0.98826, 0.98826, 0.99477)
        rgb=(0.9863, 0.9863, 0.99401)
        rgb=(0.98434, 0.98434, 0.99327)
        rgb=(0.98239, 0.98239, 0.99257)
        rgb=(0.98043, 0.98043, 0.9919)
        rgb=(0.97847, 0.97847, 0.99125)
        rgb=(0.97652, 0.97652, 0.99064)
        rgb=(0.97456, 0.97456, 0.99006)
        rgb=(0.9726, 0.9726, 0.98951)
        rgb=(0.97065, 0.97065, 0.98899)
        rgb=(0.96869, 0.96869, 0.9885)
        rgb=(0.96673, 0.96673, 0.98804)
        rgb=(0.96477, 0.96477, 0.98762)
        rgb=(0.96282, 0.96282, 0.98722)
        rgb=(0.96086, 0.96086, 0.98685)
        rgb=(0.9589, 0.9589, 0.98652)
        rgb=(0.95695, 0.95695, 0.98621)
        rgb=(0.95499, 0.95499, 0.98593)
        rgb=(0.95303, 0.95303, 0.98569)
        rgb=(0.95108, 0.95108, 0.98548)
        rgb=(0.94912, 0.94912, 0.98529)
        rgb=(0.94716, 0.94716, 0.98514)
        rgb=(0.94521, 0.94521, 0.98502)
        rgb=(0.94325, 0.94325, 0.98493)
        rgb=(0.94129, 0.94129, 0.98486)
        rgb=(0.93933, 0.93933, 0.98483)
        rgb=(0.93738, 0.93738, 0.98483)
        rgb=(0.93542, 0.93542, 0.98486)
        rgb=(0.93346, 0.93346, 0.98493)
        rgb=(0.93151, 0.93151, 0.98502)
        rgb=(0.92955, 0.92955, 0.98514)
        rgb=(0.92759, 0.92759, 0.98529)
        rgb=(0.92564, 0.92564, 0.98548)
        rgb=(0.92368, 0.92368, 0.98569)
        rgb=(0.92172, 0.92172, 0.98593)
        rgb=(0.91977, 0.91977, 0.98621)
        rgb=(0.91781, 0.91781, 0.98652)
        rgb=(0.91585, 0.91585, 0.98685)
        rgb=(0.91389, 0.91389, 0.98722)
        rgb=(0.91194, 0.91194, 0.98762)
        rgb=(0.90998, 0.90998, 0.98804)
        rgb=(0.90802, 0.90802, 0.9885)
        rgb=(0.90607, 0.90607, 0.98899)
        rgb=(0.90411, 0.90411, 0.98951)
        rgb=(0.90215, 0.90215, 0.99006)
        rgb=(0.9002, 0.9002, 0.99064)
        rgb=(0.89824, 0.89824, 0.99125)
        rgb=(0.89628, 0.89628, 0.9919)
        rgb=(0.89432, 0.89432, 0.99257)
        rgb=(0.89237, 0.89237, 0.99327)
        rgb=(0.89041, 0.89041, 0.99401)
        rgb=(0.88845, 0.88845, 0.99477)
        rgb=(0.8865, 0.8865, 0.99557)
        rgb=(0.88454, 0.88454, 0.99639)
        rgb=(0.88258, 0.88258, 0.99725)
        rgb=(0.88063, 0.88063, 0.99813)
        rgb=(0.87867, 0.87867, 0.99905)
        rgb=(0.87671, 0.87671, 1)
        rgb=(0.87476, 0.87573, 1)
        rgb=(0.8728, 0.87479, 1)
        rgb=(0.87084, 0.87387, 1)
        rgb=(0.86888, 0.87298, 1)
        rgb=(0.86693, 0.87213, 1)
        rgb=(0.86497, 0.8713, 1)
        rgb=(0.86301, 0.87051, 1)
        rgb=(0.86106, 0.86974, 1)
        rgb=(0.8591, 0.86901, 1)
        rgb=(0.85714, 0.8683, 1)
        rgb=(0.85519, 0.86763, 1)
        rgb=(0.85323, 0.86699, 1)
        rgb=(0.85127, 0.86638, 1)
        rgb=(0.84932, 0.8658, 1)
        rgb=(0.84736, 0.86525, 1)
        rgb=(0.8454, 0.86473, 1)
        rgb=(0.84344, 0.86424, 1)
        rgb=(0.84149, 0.86378, 1)
        rgb=(0.83953, 0.86335, 1)
        rgb=(0.83757, 0.86295, 1)
        rgb=(0.83562, 0.86259, 1)
        rgb=(0.83366, 0.86225, 1)
        rgb=(0.8317, 0.86194, 1)
        rgb=(0.82975, 0.86167, 1)
        rgb=(0.82779, 0.86142, 1)
        rgb=(0.82583, 0.86121, 1)
        rgb=(0.82387, 0.86103, 1)
        rgb=(0.82192, 0.86087, 1)
        rgb=(0.81996, 0.86075, 1)
        rgb=(0.818, 0.86066, 1)
        rgb=(0.81605, 0.8606, 1)
        rgb=(0.81409, 0.86057, 1)
        rgb=(0.81213, 0.86057, 1)
        rgb=(0.81018, 0.8606, 1)
        rgb=(0.80822, 0.86066, 1)
        rgb=(0.80626, 0.86075, 1)
        rgb=(0.80431, 0.86087, 1)
        rgb=(0.80235, 0.86103, 1)
        rgb=(0.80039, 0.86121, 1)
        rgb=(0.79843, 0.86142, 1)
        rgb=(0.79648, 0.86167, 1)
        rgb=(0.79452, 0.86194, 1)
        rgb=(0.79256, 0.86225, 1)
        rgb=(0.79061, 0.86259, 1)
        rgb=(0.78865, 0.86295, 1)
        rgb=(0.78669, 0.86335, 1)
        rgb=(0.78474, 0.86378, 1)
        rgb=(0.78278, 0.86424, 1)
        rgb=(0.78082, 0.86473, 1)
        rgb=(0.77886, 0.86525, 1)
        rgb=(0.77691, 0.8658, 1)
        rgb=(0.77495, 0.86638, 1)
        rgb=(0.77299, 0.86699, 1)
        rgb=(0.77104, 0.86763, 1)
        rgb=(0.76908, 0.8683, 1)
        rgb=(0.76712, 0.86901, 1)
        rgb=(0.76517, 0.86974, 1)
        rgb=(0.76321, 0.87051, 1)
        rgb=(0.76125, 0.8713, 1)
        rgb=(0.7593, 0.87213, 1)
        rgb=(0.75734, 0.87298, 1)
        rgb=(0.75538, 0.87387, 1)
        rgb=(0.75342, 0.87479, 1)
        rgb=(0.75147, 0.87573, 1)
        rgb=(0.74951, 0.87671, 1)
        rgb=(0.74755, 0.87772, 1)
        rgb=(0.7456, 0.87876, 1)
        rgb=(0.74364, 0.87983, 1)
        rgb=(0.74168, 0.88093, 1)
        rgb=(0.73973, 0.88206, 1)
        rgb=(0.73777, 0.88323, 1)
        rgb=(0.73581, 0.88442, 1)
        rgb=(0.73386, 0.88564, 1)
        rgb=(0.7319, 0.88689, 1)
        rgb=(0.72994, 0.88818, 1)
        rgb=(0.72798, 0.88949, 1)
        rgb=(0.72603, 0.89084, 1)
        rgb=(0.72407, 0.89222, 1)
        rgb=(0.72211, 0.89362, 1)
        rgb=(0.72016, 0.89506, 1)
        rgb=(0.7182, 0.89653, 1)
        rgb=(0.71624, 0.89802, 1)
        rgb=(0.71429, 0.89955, 1)
        rgb=(0.71233, 0.90111, 1)
        rgb=(0.71037, 0.9027, 1)
        rgb=(0.70841, 0.90432, 1)
        rgb=(0.70646, 0.90597, 1)
        rgb=(0.7045, 0.90766, 1)
        rgb=(0.70254, 0.90937, 1)
        rgb=(0.70059, 0.91111, 1)
        rgb=(0.69863, 0.91289, 1)
        rgb=(0.69667, 0.91469, 1)
        rgb=(0.69472, 0.91652, 1)
        rgb=(0.69276, 0.91839, 1)
        rgb=(0.6908, 0.92028, 1)
        rgb=(0.68885, 0.92221, 1)
        rgb=(0.68689, 0.92417, 1)
        rgb=(0.68493, 0.92616, 1)
        rgb=(0.68297, 0.92817, 1)
        rgb=(0.68102, 0.93022, 1)
        rgb=(0.67906, 0.9323, 1)
        rgb=(0.6771, 0.93441, 1)
        rgb=(0.67515, 0.93655, 1)
        rgb=(0.67319, 0.93872, 1)
        rgb=(0.67123, 0.94092, 1)
        rgb=(0.66928, 0.94316, 1)
        rgb=(0.66732, 0.94542, 1)
        rgb=(0.66536, 0.94771, 1)
        rgb=(0.66341, 0.95004, 1)
        rgb=(0.66145, 0.95239, 1)
        rgb=(0.65949, 0.95478, 1)
        rgb=(0.65753, 0.95719, 1)
        rgb=(0.65558, 0.95964, 1)
        rgb=(0.65362, 0.96211, 1)
        rgb=(0.65166, 0.96462, 1)
        rgb=(0.64971, 0.96716, 1)
        rgb=(0.64775, 0.96973, 1)
        rgb=(0.64579, 0.97233, 1)
        rgb=(0.64384, 0.97496, 1)
        rgb=(0.64188, 0.97762, 1)
        rgb=(0.63992, 0.98031, 1)
        rgb=(0.63796, 0.98303, 1)
        rgb=(0.63601, 0.98578, 1)
        rgb=(0.63405, 0.98856, 1)
        rgb=(0.63209, 0.99138, 1)
        rgb=(0.63014, 0.99422, 1)
        rgb=(0.62818, 0.9971, 1)
        rgb=(0.62622, 1, 1)
        rgb=(0.6272, 1, 0.99706)
        rgb=(0.62821, 1, 0.9941)
        rgb=(0.62925, 1, 0.9911)
        rgb=(0.63032, 1, 0.98807)
        rgb=(0.63142, 1, 0.98502)
        rgb=(0.63255, 1, 0.98193)
        rgb=(0.63371, 1, 0.97881)
        rgb=(0.63491, 1, 0.97566)
        rgb=(0.63613, 1, 0.97248)
        rgb=(0.63738, 1, 0.96927)
        rgb=(0.63867, 1, 0.96603)
        rgb=(0.63998, 1, 0.96276)
        rgb=(0.64133, 1, 0.95945)
        rgb=(0.6427, 1, 0.95612)
        rgb=(0.64411, 1, 0.95276)
        rgb=(0.64555, 1, 0.94936)
        rgb=(0.64702, 1, 0.94594)
        rgb=(0.64851, 1, 0.94248)
        rgb=(0.65004, 1, 0.939)
        rgb=(0.6516, 1, 0.93548)
        rgb=(0.65319, 1, 0.93193)
        rgb=(0.65481, 1, 0.92836)
        rgb=(0.65646, 1, 0.92475)
        rgb=(0.65815, 1, 0.92111)
        rgb=(0.65986, 1, 0.91744)
        rgb=(0.6616, 1, 0.91374)
        rgb=(0.66337, 1, 0.91001)
        rgb=(0.66518, 1, 0.90625)
        rgb=(0.66701, 1, 0.90246)
        rgb=(0.66888, 1, 0.89864)
        rgb=(0.67077, 1, 0.89478)
        rgb=(0.6727, 1, 0.8909)
        rgb=(0.67466, 1, 0.88699)
        rgb=(0.67665, 1, 0.88304)
        rgb=(0.67866, 1, 0.87907)
        rgb=(0.68071, 1, 0.87506)
        rgb=(0.68279, 1, 0.87102)
        rgb=(0.6849, 1, 0.86696)
        rgb=(0.68704, 1, 0.86286)
        rgb=(0.68921, 1, 0.85873)
        rgb=(0.69141, 1, 0.85457)
        rgb=(0.69365, 1, 0.85039)
        rgb=(0.69591, 1, 0.84617)
        rgb=(0.6982, 1, 0.84192)
        rgb=(0.70053, 1, 0.83763)
        rgb=(0.70288, 1, 0.83332)
        rgb=(0.70527, 1, 0.82898)
        rgb=(0.70768, 1, 0.82461)
        rgb=(0.71013, 1, 0.82021)
        rgb=(0.7126, 1, 0.81577)
        rgb=(0.71511, 1, 0.81131)
        rgb=(0.71765, 1, 0.80681)
        rgb=(0.72022, 1, 0.80229)
        rgb=(0.72282, 1, 0.79773)
        rgb=(0.72545, 1, 0.79314)
        rgb=(0.72811, 1, 0.78853)
        rgb=(0.7308, 1, 0.78388)
        rgb=(0.73352, 1, 0.7792)
        rgb=(0.73627, 1, 0.77449)
        rgb=(0.73905, 1, 0.76975)
        rgb=(0.74187, 1, 0.76498)
        rgb=(0.74471, 1, 0.76018)
        rgb=(0.74758, 1, 0.75535)
        rgb=(0.75049, 1, 0.75049)
        rgb=(0.75342, 1, 0.7456)
        rgb=(0.75639, 1, 0.74067)
        rgb=(0.75939, 1, 0.73572)
        rgb=(0.76241, 1, 0.73074)
        rgb=(0.76547, 1, 0.72572)
        rgb=(0.76856, 1, 0.72068)
        rgb=(0.77168, 1, 0.7156)
        rgb=(0.77483, 1, 0.71049)
        rgb=(0.77801, 1, 0.70536)
        rgb=(0.78122, 1, 0.70019)
        rgb=(0.78446, 1, 0.69499)
        rgb=(0.78773, 1, 0.68976)
        rgb=(0.79103, 1, 0.6845)
        rgb=(0.79437, 1, 0.67921)
        rgb=(0.79773, 1, 0.67389)
        rgb=(0.80113, 1, 0.66854)
        rgb=(0.80455, 1, 0.66316)
        rgb=(0.80801, 1, 0.65775)
        rgb=(0.81149, 1, 0.65231)
        rgb=(0.81501, 1, 0.64683)
        rgb=(0.81855, 1, 0.64133)
        rgb=(0.82213, 1, 0.63579)
        rgb=(0.82574, 1, 0.63023)
        rgb=(0.82938, 1, 0.62463)
        rgb=(0.83305, 1, 0.61901)
        rgb=(0.83675, 1, 0.61335)
        rgb=(0.84048, 1, 0.60766)
        rgb=(0.84424, 1, 0.60194)
        rgb=(0.84803, 1, 0.5962)
        rgb=(0.85185, 1, 0.59042)
        rgb=(0.85571, 1, 0.58461)
        rgb=(0.85959, 1, 0.57877)
        rgb=(0.8635, 1, 0.5729)
        rgb=(0.86745, 1, 0.56699)
        rgb=(0.87142, 1, 0.56106)
        rgb=(0.87543, 1, 0.5551)
        rgb=(0.87946, 1, 0.54911)
        rgb=(0.88353, 1, 0.54308)
        rgb=(0.88763, 1, 0.53703)
        rgb=(0.89176, 1, 0.53094)
        rgb=(0.89591, 1, 0.52483)
        rgb=(0.9001, 1, 0.51868)
        rgb=(0.90432, 1, 0.51251)
        rgb=(0.90857, 1, 0.5063)
        rgb=(0.91285, 1, 0.50006)
        rgb=(0.91717, 1, 0.49379)
        rgb=(0.92151, 1, 0.48749)
        rgb=(0.92588, 1, 0.48116)
        rgb=(0.93028, 1, 0.4748)
        rgb=(0.93472, 1, 0.46841)
        rgb=(0.93918, 1, 0.46199)
        rgb=(0.94368, 1, 0.45554)
        rgb=(0.9482, 1, 0.44906)
        rgb=(0.95276, 1, 0.44255)
        rgb=(0.95734, 1, 0.436)
        rgb=(0.96196, 1, 0.42943)
        rgb=(0.96661, 1, 0.42282)
        rgb=(0.97129, 1, 0.41619)
        rgb=(0.976, 1, 0.40952)
        rgb=(0.98074, 1, 0.40283)
        rgb=(0.98551, 1, 0.3961)
        rgb=(0.99031, 1, 0.38934)
        rgb=(0.99514, 1, 0.38255)
        rgb=(1, 1, 0.37573)
        rgb=(1, 0.99511, 0.37378)
        rgb=(1, 0.99018, 0.37182)
        rgb=(1, 0.98523, 0.36986)
        rgb=(1, 0.98025, 0.36791)
        rgb=(1, 0.97523, 0.36595)
        rgb=(1, 0.97019, 0.36399)
        rgb=(1, 0.96511, 0.36204)
        rgb=(1, 0.96, 0.36008)
        rgb=(1, 0.95487, 0.35812)
        rgb=(1, 0.9497, 0.35616)
        rgb=(1, 0.9445, 0.35421)
        rgb=(1, 0.93927, 0.35225)
        rgb=(1, 0.93401, 0.35029)
        rgb=(1, 0.92872, 0.34834)
        rgb=(1, 0.9234, 0.34638)
        rgb=(1, 0.91805, 0.34442)
        rgb=(1, 0.91267, 0.34247)
        rgb=(1, 0.90726, 0.34051)
        rgb=(1, 0.90182, 0.33855)
        rgb=(1, 0.89634, 0.33659)
        rgb=(1, 0.89084, 0.33464)
        rgb=(1, 0.8853, 0.33268)
        rgb=(1, 0.87974, 0.33072)
        rgb=(1, 0.87414, 0.32877)
        rgb=(1, 0.86852, 0.32681)
        rgb=(1, 0.86286, 0.32485)
        rgb=(1, 0.85717, 0.3229)
        rgb=(1, 0.85146, 0.32094)
        rgb=(1, 0.84571, 0.31898)
        rgb=(1, 0.83993, 0.31703)
        rgb=(1, 0.83412, 0.31507)
        rgb=(1, 0.82828, 0.31311)
        rgb=(1, 0.82241, 0.31115)
        rgb=(1, 0.81651, 0.3092)
        rgb=(1, 0.81057, 0.30724)
        rgb=(1, 0.80461, 0.30528)
        rgb=(1, 0.79862, 0.30333)
        rgb=(1, 0.79259, 0.30137)
        rgb=(1, 0.78654, 0.29941)
        rgb=(1, 0.78045, 0.29746)
        rgb=(1, 0.77434, 0.2955)
        rgb=(1, 0.76819, 0.29354)
        rgb=(1, 0.76202, 0.29159)
        rgb=(1, 0.75581, 0.28963)
        rgb=(1, 0.74957, 0.28767)
        rgb=(1, 0.7433, 0.28571)
        rgb=(1, 0.737, 0.28376)
        rgb=(1, 0.73068, 0.2818)
        rgb=(1, 0.72432, 0.27984)
        rgb=(1, 0.71792, 0.27789)
        rgb=(1, 0.7115, 0.27593)
        rgb=(1, 0.70505, 0.27397)
        rgb=(1, 0.69857, 0.27202)
        rgb=(1, 0.69206, 0.27006)
        rgb=(1, 0.68551, 0.2681)
        rgb=(1, 0.67894, 0.26614)
        rgb=(1, 0.67233, 0.26419)
        rgb=(1, 0.6657, 0.26223)
        rgb=(1, 0.65903, 0.26027)
        rgb=(1, 0.65234, 0.25832)
        rgb=(1, 0.64561, 0.25636)
        rgb=(1, 0.63885, 0.2544)
        rgb=(1, 0.63206, 0.25245)
        rgb=(1, 0.62524, 0.25049)
        rgb=(1, 0.6184, 0.24853)
        rgb=(1, 0.61152, 0.24658)
        rgb=(1, 0.6046, 0.24462)
        rgb=(1, 0.59766, 0.24266)
        rgb=(1, 0.59069, 0.2407)
        rgb=(1, 0.58369, 0.23875)
        rgb=(1, 0.57666, 0.23679)
        rgb=(1, 0.56959, 0.23483)
        rgb=(1, 0.5625, 0.23288)
        rgb=(1, 0.55538, 0.23092)
        rgb=(1, 0.54822, 0.22896)
        rgb=(1, 0.54103, 0.22701)
        rgb=(1, 0.53382, 0.22505)
        rgb=(1, 0.52657, 0.22309)
        rgb=(1, 0.51929, 0.22114)
        rgb=(1, 0.51199, 0.21918)
        rgb=(1, 0.50465, 0.21722)
        rgb=(1, 0.49728, 0.21526)
        rgb=(1, 0.48988, 0.21331)
        rgb=(1, 0.48245, 0.21135)
        rgb=(1, 0.47499, 0.20939)
        rgb=(1, 0.4675, 0.20744)
        rgb=(1, 0.45997, 0.20548)
        rgb=(1, 0.45242, 0.20352)
        rgb=(1, 0.44484, 0.20157)
        rgb=(1, 0.43722, 0.19961)
        rgb=(1, 0.42958, 0.19765)
        rgb=(1, 0.42191, 0.19569)
        rgb=(1, 0.4142, 0.19374)
        rgb=(1, 0.40646, 0.19178)
        rgb=(1, 0.3987, 0.18982)
        rgb=(1, 0.3909, 0.18787)
        rgb=(1, 0.38307, 0.18591)
        rgb=(1, 0.37521, 0.18395)
        rgb=(1, 0.36733, 0.182)
        rgb=(1, 0.35941, 0.18004)
        rgb=(1, 0.35146, 0.17808)
        rgb=(1, 0.34347, 0.17613)
        rgb=(1, 0.33546, 0.17417)
        rgb=(1, 0.32742, 0.17221)
        rgb=(1, 0.31935, 0.17025)
        rgb=(1, 0.31125, 0.1683)
        rgb=(1, 0.30311, 0.16634)
        rgb=(1, 0.29495, 0.16438)
        rgb=(1, 0.28675, 0.16243)
        rgb=(1, 0.27853, 0.16047)
        rgb=(1, 0.27027, 0.15851)
        rgb=(1, 0.26199, 0.15656)
        rgb=(1, 0.25367, 0.1546)
        rgb=(1, 0.24532, 0.15264)
        rgb=(1, 0.23694, 0.15068)
        rgb=(1, 0.22853, 0.14873)
        rgb=(1, 0.2201, 0.14677)
        rgb=(1, 0.21163, 0.14481)
        rgb=(1, 0.20312, 0.14286)
        rgb=(1, 0.19459, 0.1409)
        rgb=(1, 0.18603, 0.13894)
        rgb=(1, 0.17744, 0.13699)
        rgb=(1, 0.16882, 0.13503)
        rgb=(1, 0.16016, 0.13307)
        rgb=(1, 0.15148, 0.13112)
        rgb=(1, 0.14277, 0.12916)
        rgb=(1, 0.13402, 0.1272)
        rgb=(1, 0.12524, 0.12524)
        rgb=(0.99315, 0.12329, 0.12329)
        rgb=(0.98627, 0.12133, 0.12133)
        rgb=(0.97936, 0.11937, 0.11937)
        rgb=(0.97242, 0.11742, 0.11742)
        rgb=(0.96545, 0.11546, 0.11546)
        rgb=(0.95845, 0.1135, 0.1135)
        rgb=(0.95141, 0.11155, 0.11155)
        rgb=(0.94435, 0.10959, 0.10959)
        rgb=(0.93726, 0.10763, 0.10763)
        rgb=(0.93013, 0.10568, 0.10568)
        rgb=(0.92298, 0.10372, 0.10372)
        rgb=(0.91579, 0.10176, 0.10176)
        rgb=(0.90857, 0.099804, 0.099804)
        rgb=(0.90133, 0.097847, 0.097847)
        rgb=(0.89405, 0.09589, 0.09589)
        rgb=(0.88674, 0.093933, 0.093933)
        rgb=(0.8794, 0.091977, 0.091977)
        rgb=(0.87203, 0.09002, 0.09002)
        rgb=(0.86463, 0.088063, 0.088063)
        rgb=(0.8572, 0.086106, 0.086106)
        rgb=(0.84974, 0.084149, 0.084149)
        rgb=(0.84225, 0.082192, 0.082192)
        rgb=(0.83473, 0.080235, 0.080235)
        rgb=(0.82718, 0.078278, 0.078278)
        rgb=(0.81959, 0.076321, 0.076321)
        rgb=(0.81198, 0.074364, 0.074364)
        rgb=(0.80434, 0.072407, 0.072407)
        rgb=(0.79666, 0.07045, 0.07045)
        rgb=(0.78896, 0.068493, 0.068493)
        rgb=(0.78122, 0.066536, 0.066536)
        rgb=(0.77345, 0.064579, 0.064579)
        rgb=(0.76566, 0.062622, 0.062622)
        rgb=(0.75783, 0.060665, 0.060665)
        rgb=(0.74997, 0.058708, 0.058708)
        rgb=(0.74208, 0.056751, 0.056751)
        rgb=(0.73416, 0.054795, 0.054795)
        rgb=(0.72621, 0.052838, 0.052838)
        rgb=(0.71823, 0.050881, 0.050881)
        rgb=(0.71022, 0.048924, 0.048924)
        rgb=(0.70218, 0.046967, 0.046967)
        rgb=(0.6941, 0.04501, 0.04501)
        rgb=(0.686, 0.043053, 0.043053)
        rgb=(0.67787, 0.041096, 0.041096)
        rgb=(0.6697, 0.039139, 0.039139)
        rgb=(0.66151, 0.037182, 0.037182)
        rgb=(0.65328, 0.035225, 0.035225)
        rgb=(0.64503, 0.033268, 0.033268)
        rgb=(0.63674, 0.031311, 0.031311)
        rgb=(0.62842, 0.029354, 0.029354)
        rgb=(0.62008, 0.027397, 0.027397)
        rgb=(0.6117, 0.02544, 0.02544)
        rgb=(0.60329, 0.023483, 0.023483)
        rgb=(0.59485, 0.021526, 0.021526)
        rgb=(0.58638, 0.019569, 0.019569)
        rgb=(0.57788, 0.017613, 0.017613)
        rgb=(0.56935, 0.015656, 0.015656)
        rgb=(0.56079, 0.013699, 0.013699)
        rgb=(0.5522, 0.011742, 0.011742)
        rgb=(0.54357, 0.0097847, 0.0097847)
        rgb=(0.53492, 0.0078278, 0.0078278)
        rgb=(0.52624, 0.0058708, 0.0058708)
        rgb=(0.51752, 0.0039139, 0.0039139)
        rgb=(0.50878, 0.0019569, 0.0019569)
        rgb=(0.5, 0, 0)
    }
}

\usepackage[bookmarks=false]{hyperref}
\hypersetup{bookmarks=false}

\usepackage{subcaption}
\usepackage{makecell} 
\usepackage{lipsum} 

\usepackage{textcase}
\usepackage[tablename=TABLE,font=small]{caption}
\DeclareCaptionTextFormat{up}{\MakeTextUppercase{#1}}
\newcommand{\wrt}{w.\,r.\,t.\ }

\newcommand\blfootnote[1]{%
  \begingroup
  \renewcommand\thefootnote{}\footnote{#1}%
  \addtocounter{footnote}{-1}%
  \endgroup
}

\newtheorem{remark}{Remark}
\newtheorem*{remark*}{Remark}

\begin{document}
\begin{NoHyper}
\title{Angular Estimation Comparison with ISAC PoC}

\author{

\IEEEauthorblockN{
        Alexander~Felix\IEEEauthorrefmark{1}\IEEEauthorrefmark{2},
        Rudolf~Hoffmann\IEEEauthorrefmark{3}, 
        Marcus~Henninger\IEEEauthorrefmark{1},
        Stephan~ten~Brink\IEEEauthorrefmark{2},
        and Silvio~Mandelli\IEEEauthorrefmark{1}
        }
        
	\IEEEauthorblockA{
	\IEEEauthorrefmark{1}Nokia Bell Labs Stuttgart, 70469 Stuttgart, Germany \\
	\IEEEauthorrefmark{2}Institute of Telecommunications, University of Stuttgart, 70569 Stuttgart, Germany \\
    \IEEEauthorrefmark{3}GPP Communication GmbH \& Co. KG, 82041 Oberhaching, Germany \\
	E-mail: alexander.felix@nokia.com}        
\thanks{TBD: Further notes.}}

\maketitle

\newacronym{2D}{2D}{two-dimensional}
\newacronym{5G}{5G}{fifth generation}
\newacronym{6G}{6G}{sixth generation}
\newacronym{aal}{AAL}{array aperture  line}
\newacronym{awgn}{AWGN}{additive white Gaussian noise}
\newacronym{cfar}{CFAR}{constant false alarm rate}
\newacronym{dft}{DFT}{discrete Fourier transform}
\newacronym{doa}{DoA}{direction of arrival}
\newacronym{idft}{IDFT}{inverse discrete Fourier transform}
\newacronym{isac}{ISAC}{Integrated Sensing and Communications}
\newacronym{kpi}{KPI}{key performance indicator}
\newacronym[plural={MRAs}]{mra}{MRA}{minimum redundancy array}
\newacronym{lmf}{LMF}{Location Management Function}
\newacronym{mf}{MF}{management function}
\newacronym{mp}{MP}{matching pursuit}
\newacronym{naf}{NAF}{normalized angular frequency}
\newacronym{NRPPa}{NRPPa}{NR Positioning Protocol A}
\newacronym{ofdm}{OFDM}{orthogonal frequency-division multiplexing}
\newacronym{omp}{OMP}{orthogonal matching pursuit}
\newacronym{poc}{PoC}{proof of concept}
\newacronym{pov}{POV}{point of view}
\newacronym[plural={PSFs}]{psf}{PSF}{point spread function}
\newacronym[plural={RUs}]{ru}{RU}{radio unit}
\newacronym{rx}{RX}{receiver}
\newacronym{rmse}{RMSE}{root mean squared error}
\newacronym{sar}{SAR}{synthetic-aperture radar}
\newacronym{sara}{SARA}{Sampling and Reconstructing Angular Domains with Uniform Arrays}
\newacronym{semf}{SeMF}{Sensing Management Function}
\newacronym{tx}{TX}{transmitter}
\newacronym[plural={TRPs}]{trp}{TRP}{transmission reception point}
\newacronym[plural={ULAs}]{ula}{ULA}{uniform linear array}
\newacronym[plural={URAs}]{ura}{URA}{uniform rectangular array}

\if\isshorter1
\begin{abstract}

The introduction of \acrfull{isac} in cellular systems is not expected to result in a shift away from the popular choice of cost- and energy-efficient analog or hybrid beamforming structures. However, this comes at the cost of limiting the angular capabilities to a confined space per acquisitions. 
Thus, as a prerequisite for the successful implementation of numerous \acrshort{isac} use cases, the need for an optimal angular estimation of targets and their separation based on the minimal number of angular samples arises.

In this work, different approaches for angular estimation based on a minimal, \acrshort{dft}-based set of angular samples are evaluated. The samples are acquired through sweeping multiple beams of an \acrshort{isac} \acrfull{poc} in the industrial scenario of the \textit{ARENA2036}. 
The study's findings indicate that interpolation approaches are more effective for generalizing across different types of angular scenarios. While the \acrfull{omp} approach exhibits the most accurate estimation for a single, strong and clearly discriminable target, the \acrshort{dft}-based interpolation approach demonstrates the best overall estimation performance.

\end{abstract}

\else

\fi

\if\arxivdisclaimer1
\blfootnote{This work has been submitted to the IEEE for possible publication. Copyright may be transferred without notice, after which this version may no longer be accessible.}
\vspace{-0.50cm}
\else
\vspace{0.2cm}
\begin{IEEEkeywords}
Monostatic radar, Angular sampling and interpolation, Integrated Sensing and Communications (ISAC).
\end{IEEEkeywords}
\fi

\IEEEpeerreviewmaketitle

\glsresetall
\section{Introduction}

\gls{isac} as part of \gls{6G} cellular networks is set out to establish radar-like capabilities on top of communications systems. Depending on the type of task which needs to be completed in a time slot, the system needs to optimize for different, potentially conflicting \acrshortpl{kpi}~\cite{liuJointRadarCommunication2020, 3GPP_TR_22837}. However, it is reasonable to hypothesize that the majority of early \gls{isac} systems will prioritize communications capabilities. 
Over the last decades, MIMO and beamforming techniques have emerged as increasingly prevalent methodologies in cellular deployments, enabling spatial sampling and segmentation of their surrounding environments.
The ability of spatial imaging is affected by the type of beamforming structures in the deployment. Due to cost and energy benefits, analog or hybrid beamforming structures remain up to now a popular choice for commercial deployments.
Nevertheless, in contrast to digital beamforming, they impose constraints on the number of scannable angles in a single acquisition step~\cite{saha2021quantitative, chaudhari5GPhysicalLayer2021}.

As the \gls{6G} standardization gains momentum, we expect to see more practical evaluations of \gls{isac} systems. So far, Nokia has conducted numerous \gls{poc} measurement campaigns with various partners using a commercially available gNodeB~\cite{wild2023integrated, muthineniDeepLearningBasedData2025b}. 
Other work is predominantly based on USRP systems, e.g., a fully-digital evaluation with horn antennas~\cite{kumariJCR70LowComplexityMillimeterWave2020}, a bistatic topology oversampling the angular domain~\cite{paidimarriEyeBeamMmWave5GCompliant2024}, and a multistatic data set mainly with fixed beam acquisitions~\cite{beusterEnhancingSituationalAwareness2025}. 
However, to the best of our knowledge, this is the first work on the minimal acquisition and estimation of targets in the angular domain via beam sweeping of an \gls{isac} \gls{poc} system.

When using an \gls{isac} system with confined angular acquisition capabilities, it is crucial to extract the optimal angular response from a minimal set of angular samples combined with optimal reconstruction capabilities, preserving capabilities for the communications tasks. 
Based on coarray theory~\cite{hoctor1990unifying} and utilizing Fourier principles, methodologies solving this challenge for monostatic setups are discussed from a theoretical perspective and verified via simulations in~\cite{mandelli2022sampling, felixAntennaArrayDesign2024a}. Such a \gls{dft}-based approach is utilized in this work, providing the minimal angles to be scanned to perfectly reconstruct the angular domain of a uniform array trough \gls{dft}-based interpolation.
As an alternative approach, \gls{omp}, widely used in the radar compressive sensing domain~\cite{herman2009high}, is applied on the same minimal set of \gls{dft} samples. Adapting \gls{omp} to the analog beam sweeping application yields a sparse high-resolution angular reconstruction. 

We base our investigation and measurement campaign on a \gls{poc} with confined angular acquisition capabilities~\cite{wild2023integrated}. The setup consisting of two separate, quasi co-located, \gls{5G} NR \glspl{ru} is considered monostatic. To evaluate the setup's capabilities in the angular domain, different reflectors are placed at a fixed distance from the cellular setup while varying the distance between the reflectors.

Our contributions are the following:
\begin{itemize}
    \item Angular estimation for real-world \gls{isac} applications: We conduct a comprehensive measurement campaign within angular capability limits of a cellular FR2 setup in practical \gls{isac} scenarios, specifically within the \textit{ARENA2036} industrial environment.
    \item Evaluation of state-of-the-art angular estimation methods using minimal samples: We deliver a comparative study of leading angular estimation techniques, including the evaluation of an optimal \gls{dft}-based acquisition and interpolation method tailored for monostatic setups, as well as an \gls{omp} estimator designed to maximize accuracy and efficiency with minimal acquired samples. 
\end{itemize}

\section{DFT Sampling and Interpolation}

In order to sample the angular response of a MIMO setup, which electronically steers through beamforming operations over its antennas, an arbitrary direction is defined by the azimuth and elevation scan angle pair $(\theta_s, \phi_s)$ as
\begin{align}
\thinmuskip=1.5mu
\medmuskip=1.5mu
\thickmuskip=2mu
\mathbf{u}\left(\theta_s, \phi_s\right)= [x_s,z_s]^{\intercal} = \left[{\cos(\phi_s)}{\sin(\theta_s)}, {\sin(\phi_s)} \right]^{\intercal},
\label{eq:DirectionXZ}
\end{align}
We consider a MIMO setup consisting of a \gls{tx} with $N$ elements, and a \gls{rx} with $M$ elements, both have uniform element spacing $d$. The normalized element positions of the arrays are given as $\mathbf{p}_n^{(\cdot)}~=~\left[\sfrac{x_n}{d}, \sfrac{z_n}{d}\right]^\text{T}$, as such being the unitary antenna element domain referred to as \gls{aal}. 
The coherent beamformed signal of such a monostatic setup, depending on the scanned angle pair $(\theta_s, \phi_s)$, can be expressed as
\begin{align}
\thinmuskip=0.5mu
\medmuskip=0.5mu
\thickmuskip=1mu
\hat{a}\left(\theta_s, \phi_s\right) = 
&\sum_{n=1}^{N} \sum_{m=1}^{M} a_{n}^{(\text{t})} a_{m}^{(\text{r})} w_{n}^{(\text{t})} w_{m}^{(\text{r})} e^{-jk_0\mathbf{u}\left(\theta_s, \phi_s\right)^\intercal\left(\mathbf{p}_{n}^{(\text{t})}+\mathbf{p}_{m}^{(\text{r})}\right)} \;,
\label{eq:EstimateMonoBeamform}
\end{align}
where we define the amplitude coefficients $a_{n}^{(\text{t})}$ at the $n$-th \gls{tx} element and $a_{m}^{(\text{r})}$ at the $m$-th \gls{rx} element. The coefficients of the \gls{rx} are representing the aggregated scatterer effects of the scenario. Furthermore, the beamforming coefficient of each antenna element is represented by $w_n^{(\text{t})}$ and correspondingly $w_m^{(\text{r})}$. Henceforth, the angular domain of such an uniform antenna array is represented in \gls{naf}, where the swept horizontal component is $\ell=\frac{d}{\lambda}\sin \left( \theta_s \right)\cos \left( \phi_s \right)$.

\subsection{Sampling}
We follow the \gls{dft}-based approach of~\cite{mandelli2022sampling} for the sampling and reconstruction of the angular domain of a \gls{ura}. The Fourier duality between the \gls{aal} and the \gls{naf}, enables the utilization of the Nyquist-Shannon sampling theorem. For a physical antenna element spacing $\leq\lambda/2$ the $N$-sized uniformly spaced \gls{aal} domain translates to an $N$-sized uniform sampling space in the \gls{naf} domain.

In the context of a radar imaging system, each \gls{naf} sample corresponds to a specific beam direction, thus, a range of multiple beams yields the angular response.

\begin{remark}
As we operate with a monostatic setup, the actual number of \gls{ura} elements to be considered for the minimal \gls{naf} samples depends on a virtual structure called the sum coarray~\cite{hoctor1990unifying}. This results in $2N-1$ \gls{naf} samples for a setup of equal \gls{tx} and \gls{rx} each with $N$ antennas.
\end{remark}

\subsection{Interpolation}

The angular response of a \gls{ula} \gls{tx} \gls{rx} system to a single point scatterer is denoted as \gls{psf}. For constant beamforming coefficients over the joint, virtual antenna elements, the \textit{sum coarray}, the \gls{psf} can be described by a Dirichlet kernel. 
The reconstruction of the angular domain can be achieved by taking the convolution of the received angular sequence and the Dirichlet kernel~\cite{mandelli2022sampling}. Furthermore, in the absence of noise, an oversampled signal can be perfectly reconstructed.

\begin{remark}
In this \gls{poc} measurement campaign, no special windowing is applied to the beamforming coefficients, thus, the optimal tradeoff of resolution and sidelobes of the joint physical setup is not achieved.
\end{remark}

\section{Orthogonal Matching Pursuit}

\vspace{-2mm}
\begin{algorithm}[H]
\caption{\Acrfull{omp}} \label{alg:omp}
\begin{algorithmic}[1]
\Require Input signal $\mathbf{y} \in \mathbb{R}^n$, dictionary $\mathbf{D} \in \mathbb{R}^{n \times m}$ (columns normalized), 
maximum iterations $K_{\max}$, threshold $\varepsilon$
\Ensure Sparse coefficient vector $\mathbf{a} \in \mathbb{R}^m$
\State Initialize residual $\mathbf{R} \gets \mathbf{y}$, support set $\mathcal{S} \gets \emptyset$, coefficients $a \gets 0_m$
\For{$k = 1$ \textbf{to} $K_{\max}$}
    \If{$\|\mathbf{R}\|_2 \le \varepsilon \|\mathbf{y}\|_2$}
        \State \textbf{break}
    \EndIf
    \State Compute correlations $\mathbf{p} \gets \mathbf{D}^H \mathbf{R}$
    \State Select index $i^* \gets \arg\max_i |p_i|$
    \State Update support $\mathcal{S} \gets \mathcal{S} \cup \{i^*\}$
    \State Extract subdictionary $\mathbf{D}_{\mathcal{S}}$
    \State Solve least squares: $\mathbf{a}_{\mathcal{S}} \gets (\mathbf{D}_{\mathcal{S}}^H \mathbf{D}_{\mathcal{S}})^{-1} \mathbf{D}_{\mathcal{S}}^H \mathbf{y}$
    \State Update residual $\mathbf{R} \gets \mathbf{y} - \mathbf{D}_{\mathcal{S}} \mathbf{a}_{\mathcal{S}}$
\EndFor
\State Assign $\mathbf{a}[\mathcal{S}] \gets \mathbf{a}_{\mathcal{S}}$
\State \Return $\mathbf{a}$
\end{algorithmic}
\end{algorithm}
\vspace{-1mm}

\Acrfull{omp} is a greedy sparse reconstruction algorithm that iteratively approximates a signal by selecting the most correlated atoms from a predefined dictionary~\cite{pati1993orthogonal, tropp2007signal}. 
At each iteration, the \gls{omp} identifies the dictionary atoms that best match the current residual and refines the estimate by solving a least-squares problem over the selected atoms, ensuring that the residual remains orthogonal to the subspace spanned by these atoms. 

Algorithm \ref{alg:omp} describes the implemented \gls{omp} tailored to the proposed application.
Here the dictionary $\mathbf{D}$ consists of normalized model beam response vectors corresponding to a discrete set of potential directions in the \gls{naf} domain.
The measurement vector $\mathbf{y}$ represents the beam response measurements obtained from the analog beam scan. 
\gls{omp} iteratively estimates the sparse coefficient vector $\mathbf{a}$ that reconstructs $\mathbf{y}$ as
\begin{equation}
    \mathbf{y} = \mathbf{D}\mathbf{a} + \mathbf{n},
\end{equation}
where $\mathbf{n}$ denotes measurement noise. 
The resulting sparse vector $\mathbf{a}$ contains non-zero entries at angular positions corresponding to the dominant scatterers.

In conventional MIMO radar and array processing, \gls{omp} is typically applied to coherent complex-valued array data~\cite{herman2009high}. 
However, in the analog beamforming setup considered in this work, phase coherency over the aggregation of sequentially steered beams is not guaranteed. Therefore, only magnitude responses are considered.
These measurements can be interpreted as nonlinear projections of the spatial scene onto the beam patterns~\cite{zang2019multibeam}. 
By organizing the magnitude responses in $\mathbf{y}$ and constructing $\mathbf{D}$ from the known beam patterns, \gls{omp} can be used to reconstruct a sparse angular profile even without complete complex data.

This approach extends the principle of sparse channel or scene estimation from hybrid beamforming systems~\cite{alkhateeb2014channel, elbir2023nbaomp} to the case of analog-only beam scanning. 
The resulting sparse angular reconstruction provides high-resolution \acrfull{doa} estimates with minimal hardware complexity, aligning with current trends in \gls{isac} systems~\cite{liu2022integrated, rahman2020joint}.
\section{Experimental Setup}

Our \gls{poc} setup operates within the industrial research factory \textit{ARENA2036}. We focus in this work on evaluating the angular estimation capabilities of a cellular \gls{isac} system in FR2 based on state-of-the-art \gls{5G} NR \glspl{ru}. As illustrated in \ref{fig:poc_schematic}, the quasi co-located \glspl{ru} are mounted on a pole with a distance of approx. $1$ m between their center points and are considered monostatic. Further details on the setup can be found in~Tab.~\ref{tab:poc_params} and~\cite{wild2023integrated}.

\renewcommand{\arraystretch}{1.30} 
\begin{table}
    \centering
    \caption{\gls{poc} parameters and assumptions}
    \label{tab:poc_params}
     \begin{tabular}{|c|c|}
        \hline
        \textbf{Parameter} & \textbf{Value / Description} \\
        \Xhline{3\arrayrulewidth}
        Carrier frequency $f_c$ & 27.6 GHz\\
        \hline
        TX/RX array (horizontal) & \glspl{ura} with $N^{(\cdot)}=8$\\
        \hline
        TX/RX element spacing (horizontal) & $\lambda/2$ \\
        \hline
        TX/RX element height & $5.12$ m and $4.14$ m \\
        \hline
        Distance between \glspl{ru} and targets $r$  & 18.0 m \\
        \hline
        Elevation angle  & \SI{-3.9}{^{\circ}} \\
        \hline
        Desired $P^\text{(FA)}$ of CFAR & \(10^{-6}\) \\
        \hline
    \end{tabular}
\vspace{-4mm}
\end{table}

\subsection{Angular acquisition}

The angular domain is scanned by a beam sweep from \SI{-33}{^{\circ}} to \SI{+33}{^{\circ}} with an oversampling factor of $10$, resulting in $81$ distinct \gls{naf} acquisitions. According to the \gls{dft} criterion and the angular limits of our scenario, the minimal samples are taken from the oversampled set amounting to $9$ \gls{naf} samples. In this configuration, each beam is held for a total of $6$ consecutive radio frames, with each frame recording for a duration of \SI{10}{ms}. Therefore, a full minimal sweep takes \SI{540}{ms} as opposed to \SI{4860}{ms} for the oversampled case. 
A total of $24$ frames per beam direction were recorded during this measurement campaign, which is equal to four full beam sweeps. However, for the estimation of the target positions the processed frames are limited to a single beam sweep. Hence, for the visual and numerical evaluation $6$ consecutive frames are averaged and combined. Due to the long recording time of a single beam sweep, phase coherency is not guaranteed. Thus, the magnitude of the signal is used for further processing steps.

For each radio frame, a range-Doppler periodogram is computed from its CSI utilizing Fourier transformations according to~\cite{braun2014ofdm}. Since this work focuses on angular performance, no moving objects are observed, and the Doppler domain is collapsed by selecting the zero component. Furthermore, for the numerical evaluation the range domain is limited with a priori knowledge of the rear wall area. Additionally, the range domain is collapsed by selecting the bin of the remaining range with the maximum intensity. 
Consequently, a single value is extracted for each radio frame acquisition of a beam direction.

\subsection{Angular resolution}

For the purposes of our azimuth evaluation, the parameters that are relevant are primarily those related to the horizontal dimension. 
In the \gls{naf} domain the angular resolution is $\rho=1/(2N^{(\text{1D})}-1)$, under monostatic sum coarray assumptions, where $N^{(\text{1D})}$ is the number of elements in the corresponding spatial direction. Optimally, for a joint beamforming operation, the setup containing $8$ horizontal elements should achieve a \gls{naf} resolution of $\rho^{\text{NAF}}=0.067$.
Since this \gls{poc} setup does not use ideal coefficients, only a lower resolution can be achieved.

\subsection{Scenarios}

The scenarios under test consists of different pairs of radar reflectors with varying inter-object spacing $\Delta d$. The following types of reflectors are used in the scenarios under test: two octahedral radar reflectors of dimensions $0.2\times0.2\times$ \SI{0.3}{m}, as in Fig.~\ref{fig:reflector} and from the gNodeB's \gls{pov} in Fig.~\ref{fig:scenario_reflectors},
and two acoustic walls of dimensions $1.00\times$\SI{1.80}{m}, shown in Fig.~\ref{fig:scenario_walls}.
The two chosen reflectors are positioned at a predefined distance $r$ to the base of the cellular setup, see Fig.~\ref{fig:poc_schematic}. In four increments the perpendicular inter-object spacings (\textbf{not} positions) between the two reflectors is varied: \textit{far} (\SI{7.79}{m} $\sim$ $0.209$ \gls{naf} $\sim$ \SI{25}{^{\circ}}), \textit{mid} (\SI{6.25}{m} $\sim$ $0.168$ \gls{naf} $\sim$ \SI{20}{^{\circ}}), \textit{near} (\SI{4.70}{m} $\sim$ $0.126$ \gls{naf} $\sim$ \SI{15}{^{\circ}}), \textit{resolution-limited} (\SI{3.14}{m} $\sim$ $0.084$ \gls{naf} $\sim$ \SI{10}{^{\circ}}).

\begin{figure}[t]
    \begin{subfigure}[t]{\columnwidth}
        \centering
        \resizebox{\columnwidth}{!}{
        \tikzsetnextfilename{poc_w_reflectors}

\usetikzlibrary{calc}

\def \globalscale {1.000000}
\begin{tikzpicture}[y=1pt, x=1pt, yscale=\globalscale,xscale=\globalscale, every node/.append style={scale=2.5}, inner sep=0pt, outer sep=0pt]

  \path[draw=white,fill=lightgray!40,line width=2.2pt,miter limit=10.0,cm={ 
  0.0,-1.0,1.0,0.0,(198.4, 416.6)}] (200.6, 415.9).. controls (200.6, 77.0) and 
  (248.5, -197.6) .. (307.5, -197.6).. controls (366.5, -197.6) and (414.4, 
  77.0) .. (414.4, 415.9) -- cycle;

\def \ruBaseX {558.0}
\def \ruBaseY {107.4}
\def \ruBaseWidthX {45.0}
\def \ruBaseWidthY {6.6}
\def \ruBaseWidthZ {7.4}

\path[draw=black,fill=white,line width=2.2pt,miter limit=10.0,cm={ 
-1.0,-0.0,-0.0,1.0,(1116.0, 0.0)}] 
({\ruBaseX+(\ruBaseWidthX*0.5-3.7)}, {\ruBaseY+(\ruBaseWidthZ*0.5)}) -- 
({\ruBaseX+(\ruBaseWidthX*0.5+3.7)}, {\ruBaseY-(\ruBaseWidthZ*0.5))}) -- 
({\ruBaseX+(\ruBaseWidthX*0.5+3.7)}, {\ruBaseY-(\ruBaseWidthY+\ruBaseWidthZ*0.5)}) -- 
({\ruBaseX-(\ruBaseWidthX*0.5-3.7)}, {\ruBaseY-(\ruBaseWidthY+\ruBaseWidthZ*0.5)}) -- 
({\ruBaseX-(\ruBaseWidthX*0.5+3.7)}, {\ruBaseY+(\ruBaseWidthZ*0.5-\ruBaseWidthY)}) -- 
({\ruBaseX-(\ruBaseWidthX*0.5+3.7)}, {\ruBaseY+(\ruBaseWidthZ*0.5)}) -- cycle;

\path[fill=black,fill opacity=0.1,cm={ -1.0,-0.0,-0.0,1.0,(1116.0, 0.0)}] 
({\ruBaseX-(\ruBaseWidthX*0.5+3.7)}, {\ruBaseY+(\ruBaseWidthZ*0.5)}) -- 
({\ruBaseX+(\ruBaseWidthX*0.5-3.7)}, {\ruBaseY+(\ruBaseWidthZ*0.5)}) -- 
({\ruBaseX+(\ruBaseWidthX*0.5+3.7)}, {\ruBaseY-(\ruBaseWidthZ*0.5)}) -- 
({\ruBaseX-(\ruBaseWidthX*0.5-3.7)}, {\ruBaseY-(\ruBaseWidthZ*0.5)}) -- cycle;

\path[fill=black,fill opacity=0.1,cm={ -1.0,-0.0,-0.0,1.0,(1116.0, 0.0)}] 
({\ruBaseX-(\ruBaseWidthX*0.5+3.7)}, {\ruBaseY+(\ruBaseWidthZ*0.5)}) -- 
({\ruBaseX-(\ruBaseWidthX*0.5-3.7)}, {\ruBaseY-(\ruBaseWidthZ*0.5)}) -- 
({\ruBaseX-(\ruBaseWidthX*0.5-3.7)}, {\ruBaseY-(\ruBaseWidthY+\ruBaseWidthZ*0.5)}) -- 
({\ruBaseX-(\ruBaseWidthX*0.5+3.7)}, {\ruBaseY+(\ruBaseWidthZ*0.5-\ruBaseWidthY)}) -- cycle;

\path[draw=black,line width=2.2pt,miter limit=10.0,cm={ 
-1.0,-0.0,-0.0,1.0,(1116.0, 0.0)}] 
({\ruBaseX-(\ruBaseWidthX*0.5-3.7)}, {\ruBaseY-(\ruBaseWidthY+\ruBaseWidthZ*0.5)}) -- 
({\ruBaseX-(\ruBaseWidthX*0.5-3.7)}, {\ruBaseY-(\ruBaseWidthZ*0.5)}) -- 
({\ruBaseX-(\ruBaseWidthX*0.5+3.7)}, {\ruBaseY+(\ruBaseWidthZ*0.5)}) 
({\ruBaseX-(\ruBaseWidthX*0.5-3.7)}, {\ruBaseY-(\ruBaseWidthZ*0.5)}) -- 
({\ruBaseX+(\ruBaseWidthX*0.5+3.7)}, {\ruBaseY-(\ruBaseWidthZ*0.5)});

\path[draw=black,fill=white,line width=2.2pt,miter limit=10.0] 
({\ruBaseX-5.2}, {\ruBaseY+286.1}) .. controls 
({\ruBaseX-5.2}, {\ruBaseY+287.1}) and 
({\ruBaseX-2.7}, {\ruBaseY+287.8}) .. 
({\ruBaseX+0.4}, {\ruBaseY+287.8}) .. controls 
({\ruBaseX+1.9}, {\ruBaseY+287.8}) and 
({\ruBaseX+3.3}, {\ruBaseY+287.7}) .. 
({\ruBaseX+4.4}, {\ruBaseY+287.3}) .. controls 
({\ruBaseX+5.4}, {\ruBaseY+287.0}) and 
({\ruBaseX+6.0}, {\ruBaseY+286.6}) .. 
({\ruBaseX+6.0}, {\ruBaseY+286.1}) -- 
({\ruBaseX+6.0}, {\ruBaseY+1.8}) .. controls 
({\ruBaseX+6.0}, {\ruBaseY+0.8}) and 
({\ruBaseX+3.5}, {\ruBaseY}) .. 
({\ruBaseX+0.4}, {\ruBaseY}) .. controls 
({\ruBaseX-2.7}, {\ruBaseY}) and 
({\ruBaseX-5.2}, {\ruBaseY+0.8}) .. 
({\ruBaseX-5.2}, {\ruBaseY+1.8}) -- cycle;

\path[draw=black,line width=2.2pt,miter limit=10.0] 
({\ruBaseX+6.0}, {\ruBaseY+286.1}) .. controls 
({\ruBaseX+6.0}, {\ruBaseY+285.1}) and 
({\ruBaseX+3.5}, {\ruBaseY+284.3}) .. 
({\ruBaseX+0.4}, {\ruBaseY+284.3}) .. controls 
({\ruBaseX-2.7}, {\ruBaseY+284.3}) and 
({\ruBaseX-5.2}, {\ruBaseY+285.1}) .. 
({\ruBaseX-5.2}, {\ruBaseY+286.1});

\def \reflCenterX {130.0}
\def \reflCenterY {90.0}
\def \reflWidthX {60.0}
\def \reflWidthY {60.0}
  \path[draw=black,fill=white,line width=2.2pt,miter limit=10.0] (\reflCenterX, {\reflCenterY+0.5*\reflWidthY}) 
  -- ({\reflCenterX+0.5*\reflWidthX}, \reflCenterY) -- (\reflCenterX, {\reflCenterY-0.5*\reflWidthY}) -- ({\reflCenterX-0.5*\reflWidthX}, \reflCenterY) -- cycle;

  \path[draw=black,line width=2.2pt,miter limit=10.0] ({\reflCenterX-0.5*\reflWidthX}, \reflCenterY) -- ({\reflCenterX+0.5*\reflWidthX}, 
  \reflCenterY);

  \path[draw=black,line width=2.2pt,miter limit=10.0] (\reflCenterX, ({\reflCenterY-0.5*\reflWidthY}) -- (\reflCenterX, 
  {\reflCenterY+0.5*\reflWidthY});

  \node[black, anchor=south, align=center] at (\reflCenterX,{\reflCenterY+\reflWidthY*0.7}) {T1};

\def \reflCenterUpperX {260.0}
\def \reflCenterUpperY {190.0}

  \path[draw=black,fill=white,line width=2.2pt,miter limit=10.0] (\reflCenterUpperX, {\reflCenterUpperY+0.5*\reflWidthY}) 
  -- ({\reflCenterUpperX+0.5*\reflWidthX}, \reflCenterUpperY) -- (\reflCenterUpperX, {\reflCenterUpperY-0.5*\reflWidthY}) -- ({\reflCenterUpperX-0.5*\reflWidthX}, \reflCenterUpperY) -- cycle;

  \path[draw=black,line width=2.2pt,miter limit=10.0] ({\reflCenterUpperX-0.5*\reflWidthX}, \reflCenterUpperY) -- ({\reflCenterUpperX+0.5*\reflWidthX}, 
  \reflCenterUpperY);

  \path[draw=black,line width=2.2pt,miter limit=10.0] (\reflCenterUpperX, ({\reflCenterUpperY-0.5*\reflWidthY}) -- (\reflCenterUpperX, 
  {\reflCenterUpperY+0.5*\reflWidthY});

  \node[black, anchor=south, align=center] at (\reflCenterUpperX,{\reflCenterUpperY+\reflWidthY*0.7}) {T2};

\def \scatDeltaStartX {{\reflCenterX+0.3*\reflWidthX}}
\def \scatDeltaStartY {{(\reflCenterY+0.3*\reflWidthY)}}
\def \scatDeltaEndX {{\reflCenterUpperX-0.3*\reflWidthX}}
\def \scatDeltaEndY {{(\reflCenterUpperY-0.3*\reflWidthY)}}

\draw [line width=2.2pt, arrows = {Stealth[length=13pt, inset=2pt]-Stealth[length=13pt, inset=2pt]}] (\scatDeltaStartX,\scatDeltaStartY) -- node[pos=0.4, above=5mm] {$\Delta d$} (\scatDeltaEndX,\scatDeltaEndY);

  \path[fill=orange] (535.5, 303.0) ellipse (3.8pt and 3.8pt);

  \path[draw=black,fill=white,line cap=round,line join=round,line width=4.5pt] 
  (540.5, 353.3) -- (557.5, 359.5);
  \path[draw=black,fill=white,line width=2.2pt,miter limit=10.0,cm={ 
  -0.0,-1.0,-1.0,0.0,(889.2, 889.2)}] (493.1, 359.6) -- (569.6, 359.6) -- 
  (580.9, 348.4) -- (580.9, 344.8) -- (504.4, 344.8) -- (493.1, 356.1) -- 
  (493.1, 359.6) -- cycle;

  \path[fill=black,fill opacity=0.1,cm={ -0.0,-1.0,-1.0,0.0,(889.2, 889.2)}] 
  (493.1, 359.6) -- (569.6, 359.6) -- (580.9, 348.4) -- (504.4, 348.4) -- cycle;

  \path[fill=black,fill opacity=0.1,cm={ -0.0,-1.0,-1.0,0.0,(889.2, 889.2)}] 
  (493.1, 359.6) -- (504.4, 348.4) -- (504.4, 344.8) -- (493.1, 356.1) -- cycle;

  \path[draw=black,line width=2.2pt,miter limit=10.0,cm={ 
  -0.0,-1.0,-1.0,0.0,(889.2, 889.2)}] (504.4, 344.8) -- (504.4, 348.4) -- 
  (493.1, 359.6)(504.4, 348.4) -- (580.9, 348.4);

  \path[draw=black,fill=white,line cap=round,line join=round,line width=4.5pt] 
  (540.3, 255.5) -- (557.2, 261.7);
  \path[draw=black,fill=white,line width=2.2pt,miter limit=10.0,cm={ 
  -0.0,-1.0,-1.0,0.0,(791.7, 791.7)}] (493.1, 262.1) -- (569.6, 262.1) -- 
  (580.9, 250.9) -- (580.9, 247.3) -- (504.4, 247.3) -- (493.1, 258.6) -- 
  (493.1, 262.1) -- cycle;
  
  \path[fill=black,fill opacity=0.1,cm={ -0.0,-1.0,-1.0,0.0,(791.7, 791.7)}] 
  (493.1, 262.1) -- (569.6, 262.1) -- (580.9, 250.9) -- (504.4, 250.9) -- cycle;

  \path[fill=black,fill opacity=0.1,cm={ -0.0,-1.0,-1.0,0.0,(791.7, 791.7)}] 
  (493.1, 262.1) -- (504.4, 250.9) -- (504.4, 247.3) -- (493.1, 258.6) -- cycle;

  \path[draw=black,line width=2.2pt,miter limit=10.0,cm={ 
  -0.0,-1.0,-1.0,0.0,(791.7, 791.7)}] (504.4, 247.3) -- (504.4, 250.9) -- 
  (493.1, 262.1)(504.4, 250.9) -- (580.9, 250.9);

\def \scanStartX {{\reflCenterX+0.45*\reflWidthX}}
\def \scanStartY {{\reflCenterY+0.15*\reflWidthY}}
\def \scanEndX {528.0}
\def \scanEndY {300.0}

\draw [color=orange,line width=2.2pt, arrows = {Stealth[length=13pt, inset=2pt]-Stealth[length=13pt, inset=2pt]}] (\scanStartX,\scanStartY) -- node[pos=0.6, above=8mm] {$\mathbf{u}(\theta, \phi)$} (\scanEndX,\scanEndY);

\def \radiusLowerStartX {{\ruBaseX-4}}
\def \radiusLowerStartY {{(\ruBaseY+\reflWidthY*0.5)}}
\def \radiusLowerEndX {{\reflCenterX+0.5*\reflWidthX+2}}
\def \radiusLowerEndY {\reflCenterY}
  
  \draw [line width=2.2pt, arrows = {-Stealth[length=13pt, inset=2pt]}] (\radiusLowerStartX,\radiusLowerStartY) -- node[below=2mm] {$r$} (\radiusLowerEndX,\radiusLowerEndY);

\def \radiusUpperStartX {{\ruBaseX-4}}
\def \radiusUpperStartY {{(\ruBaseY+\reflWidthY*0.5)}}
\def \radiusUpperEndX {{\reflCenterUpperX+0.5*\reflWidthX+2}}
\def \radiusUpperEndY {\reflCenterUpperY}

  \draw [line width=2.2pt, arrows = {-Stealth[length=13pt, inset=2pt]}] (\radiusUpperStartX,\radiusUpperStartY) -- node[above=2mm] {$r$} (\radiusUpperEndX,\radiusUpperEndY);

\end{tikzpicture}}
        \caption{Monostatic \gls{poc} setup of almost co-located \glspl{ru} sweeping a scenario with two radar reflectors at a fixed radius.}
        \label{fig:poc_schematic}
    \end{subfigure}
    
    \vspace{3mm}
    
    \begin{subfigure}[t]{0.33\columnwidth}
        \centering
        \input{Figures/pov_arena_reflectors}
        \caption{\gls{pov} of \gls{poc}: scenario of octahedral reflectors.}
        \label{fig:scenario_reflectors}
    \end{subfigure}%
    ~
    \begin{subfigure}[t]{0.33\columnwidth}
      \centering
      \includegraphics[width=\columnwidth]{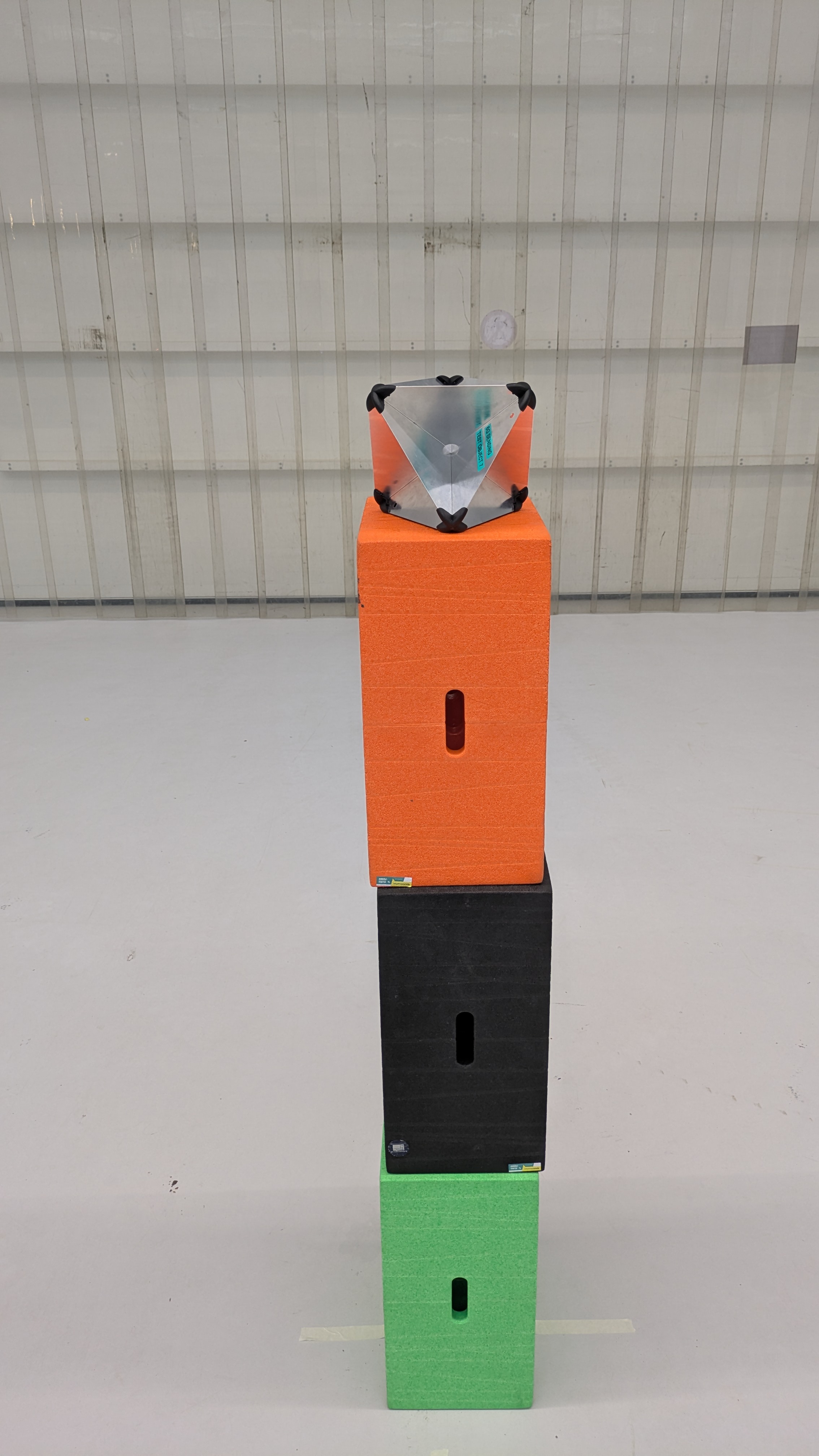}
      \caption{Octahedral radar reflector at \SI{1.5}{m} height.}
      \label{fig:reflector}
    \end{subfigure}%
    ~
    \begin{subfigure}[t]{0.33\columnwidth}
      \centering
      \input{Figures/pov_arena_walls}
      \caption{\gls{pov} of \gls{poc}: scenario of wall reflectors.}
      \label{fig:scenario_walls}
    \end{subfigure}
    \caption{Measurement scenario setup in the \textit{ARENA2036}, and overview of used radar reflectors.}
\end{figure}

\begin{figure*}[ht]
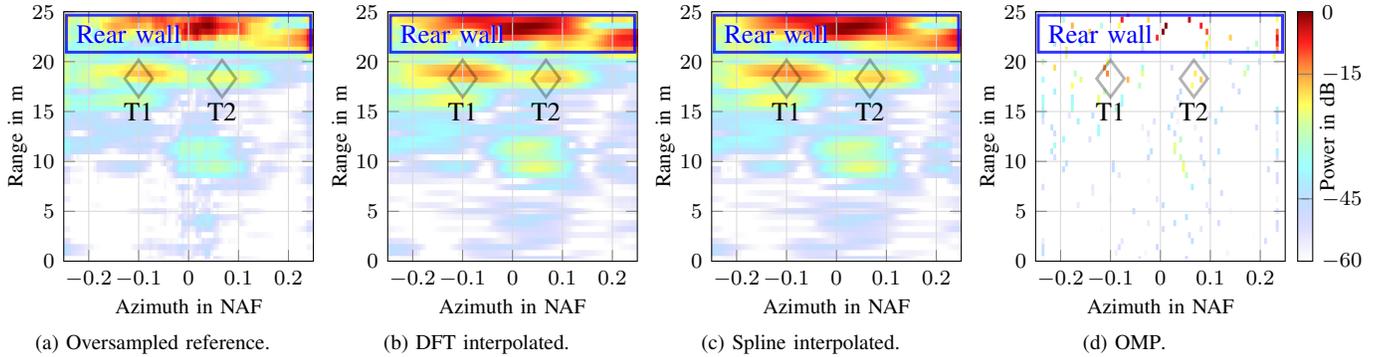

    \def \showMarker {0}
    \def \showObjects {1}
    \def \selectDist {470}
    \def \selectFrames {6frames/}
    \def \selectBDir {Data/RangeAngle/}
    \def \selectScenario {angtest_sm_reflectors_}
    \def\subfigwidth{0.225} 
    \def\figwidth{4.9cm}
    \def\figheight{\figwidth}
    \def\genvspace{-5.0mm}
    \def\genhspace{-0.2mm}  
    \centering
    \hspace*{-12.5mm}
    \begin{subfigure}[t]{\subfigwidth\textwidth}  
        \centering
        \def \selectSuffix {cm_Oversampled.png}
        \pgfplotsset{scaled y ticks=false}

\begin{tikzpicture}

    \begin{axis}[
        axis on top,
        width=\figwidth,
        height=\figheight,
        grid=major,
        grid style={solid, black!15},
        enlargelimits=false,
        xmin=-0.25, xmax=0.25,
        xlabel near ticks,
        x label style={yshift=0.2em},
        ymin=0, ymax=25,
        ytick={0,5,10,15,20,25},
        xtick={-0.3,-0.2,-0.1,0,0.1,0.2,0.3},
        ylabel near ticks,
        xlabel={Azimuth in \gls{naf}},
        ylabel={Range in m},
        y label style={yshift=-0.5em},
        label style={font=\footnotesize},
        tick label style={font=\footnotesize},
        legend cell align={left},
        legend columns = 3,
        legend style={
          fill opacity=1.0,
          draw opacity=1,
          text opacity=1,
          at={(0.999,0.999)},  
          anchor=north east,
          draw=lightgray,
          font=\scriptsize
        },
        ]

        \addplot[forget plot] graphics[
          xmin=-0.2667, 
          xmax=0.2667, 
          ymin=0, 
          ymax=25,
          includegraphics={
            trim=1 0 0 1,
            clip,}
            ] {\selectBDir\selectFrames\selectScenario\selectDist\selectSuffix};

        \def \selectPeakDir {Data/PeakEstimates/}
        \def \selectCSV {cm_peak_estimates.txt}
        \pgfplotstableread[col sep=comma]{\selectPeakDir\selectFrames\selectScenario\selectDist\selectCSV}\datatable
        
        \if\showObjects1
            \input{Figures/plot_gt}
        \fi
    
        \if\showMarker1
            \input{Figures/plot_peak_data}
        \fi
        

    \end{axis}

\end{tikzpicture}
        \vspace*{\genvspace}
        \caption{Oversampled reference.}
        \label{fig:oversampl}
    \end{subfigure}
    \hspace*{\genhspace}
    \begin{subfigure}[t]{\subfigwidth\textwidth}  
        \centering
        \def \selectSuffix {cm_DFT_Interpolation.png}
        \pgfplotsset{scaled y ticks=false}

\begin{tikzpicture}

    \begin{axis}[
        axis on top,
        width=\figwidth,
        height=\figheight,
        grid=major,
        grid style={solid, black!15},
        enlargelimits=false,
        xmin=-0.25, xmax=0.25,
        xlabel near ticks,
        x label style={yshift=0.2em},
        ymin=0, ymax=25,
        ytick={0,5,10,15,20,25},
        xtick={-0.3,-0.2,-0.1,0,0.1,0.2,0.3},
        ylabel near ticks,
        xlabel={Azimuth in \gls{naf}},
        ylabel={Range in m},
        y label style={yshift=-0.5em},
        label style={font=\footnotesize},
        tick label style={font=\footnotesize},
        legend cell align={left},
        legend columns = 3,
        legend style={
          fill opacity=1.0,
          draw opacity=1,
          text opacity=1,
          at={(0.999,0.999)},  
          anchor=north east,
          draw=lightgray,
          font=\scriptsize
        },
        ]

        \addplot[forget plot] graphics[
          xmin=-0.2667, 
          xmax=0.2667, 
          ymin=0, 
          ymax=25,
          includegraphics={
            trim=1 0 0 1,
            clip,}
            ] {\selectBDir\selectFrames\selectScenario\selectDist\selectSuffix};

        \def \selectPeakDir {Data/PeakEstimates/}
        \def \selectCSV {cm_peak_estimates.txt}
        \pgfplotstableread[col sep=comma]{\selectPeakDir\selectFrames\selectScenario\selectDist\selectCSV}\datatable
        
        \if\showObjects1
            \input{Figures/plot_gt}
        \fi
    
        \if\showMarker1
            \input{Figures/plot_peak_data}
        \fi
        

    \end{axis}

\end{tikzpicture}
        \vspace*{\genvspace}
        \caption{\gls{dft} interpolated.}
        \label{fig:dftinterp}
    \end{subfigure}
    \hspace*{\genhspace}
    \begin{subfigure}[t]{\subfigwidth\textwidth}  
        \centering
        \def \selectSuffix {cm_Spline.png}
        \pgfplotsset{scaled y ticks=false}

\begin{tikzpicture}

    \begin{axis}[
        axis on top,
        width=\figwidth,
        height=\figheight,
        grid=major,
        grid style={solid, black!15},
        enlargelimits=false,
        xmin=-0.25, xmax=0.25,
        xlabel near ticks,
        x label style={yshift=0.2em},
        ymin=0, ymax=25,
        ytick={0,5,10,15,20,25},
        xtick={-0.3,-0.2,-0.1,0,0.1,0.2,0.3},
        ylabel near ticks,
        xlabel={Azimuth in \gls{naf}},
        ylabel={Range in m},
        y label style={yshift=-0.5em},
        label style={font=\footnotesize},
        tick label style={font=\footnotesize},
        legend cell align={left},
        legend columns = 3,
        legend style={
          fill opacity=1.0,
          draw opacity=1,
          text opacity=1,
          at={(0.999,0.999)},  
          anchor=north east,
          draw=lightgray,
          font=\scriptsize
        },
        ]

        \addplot[forget plot] graphics[
          xmin=-0.2667, 
          xmax=0.2667, 
          ymin=0, 
          ymax=25,
          includegraphics={
            trim=1 0 0 1,
            clip,}
            ] {\selectBDir\selectFrames\selectScenario\selectDist\selectSuffix};

        \def \selectPeakDir {Data/PeakEstimates/}
        \def \selectCSV {cm_peak_estimates.txt}
        \pgfplotstableread[col sep=comma]{\selectPeakDir\selectFrames\selectScenario\selectDist\selectCSV}\datatable
        
        \if\showObjects1
            \input{Figures/plot_gt}
        \fi
    
        \if\showMarker1
            \input{Figures/plot_peak_data}
        \fi
        

    \end{axis}

\end{tikzpicture}
        \vspace*{\genvspace}
        \caption{Spline interpolated.}
        \label{fig:splineinterp}
    \end{subfigure}
    \hspace*{\genhspace}
    \begin{subfigure}[t]{\subfigwidth\textwidth}  
        \centering
        \def \selectSuffix {cm_OMP.png}
        \pgfplotsset{scaled y ticks=false}

\begin{tikzpicture}

    \begin{axis}[
        axis on top,
        width=\figwidth,
        height=\figheight,
        grid=major,
        grid style={solid, black!15},
        enlargelimits=false,
        xmin=-0.25, xmax=0.25,
        xlabel near ticks,
        x label style={yshift=0.2em},
        ymin=0, ymax=25,
        ytick={0,5,10,15,20,25},
        xtick={-0.3,-0.2,-0.1,0,0.1,0.2,0.3},
        ylabel near ticks,
        xlabel={Azimuth in \gls{naf}},
        ylabel={Range in m},
        y label style={yshift=-0.5em},
        label style={font=\footnotesize},
        tick label style={font=\footnotesize},
        legend cell align={left},
        legend columns = 3,
        legend style={
          fill opacity=1.0,
          draw opacity=1,
          text opacity=1,
          at={(0.999,0.999)},  
          anchor=north east,
          draw=lightgray,
          font=\scriptsize
        },
        colorbar,
        point meta min=-60,
        point meta max=0,
        colormap name=jet_inue,
        colorbar style=
            {ylabel={Power in dB}, 
            at={(1.08, 0)},
            anchor = south,
            width = 0.2cm,
            ytick={-60,-45,-15, 0},
            ylabel style={yshift=0.8cm},
            every axis/.append style=
                {font=\footnotesize}
            }
        ]

        \addplot[forget plot] graphics[
          xmin=-0.2667, 
          xmax=0.2667, 
          ymin=0, 
          ymax=25,
          includegraphics={
            trim=1 0 0 1,
            clip,}
            ] {\selectBDir\selectFrames\selectScenario\selectDist\selectSuffix};

        \def \selectPeakDir {Data/PeakEstimates/}
        \def \selectCSV {cm_peak_estimates.txt}
        \pgfplotstableread[col sep=comma]{\selectPeakDir\selectFrames\selectScenario\selectDist\selectCSV}\datatable
        
        \if\showObjects1
            \input{Figures/plot_gt}
        \fi
    
        \if\showMarker1
            \input{Figures/plot_peak_data}
        \fi
    

    \end{axis}

\end{tikzpicture}
        \vspace*{\genvspace}
        \caption{\gls{omp}.}
        \label{fig:omp}
    \end{subfigure}
    \caption {Intensity plots of two octahedral radar reflectors at a range of \SI{18}{m} with an inter-object spacing of \SI{4.70}{m} $\sim$ \SI{15}{^{\circ}} in azimuth. Four angular acquisition techniques are compared, each consisting of $81$ angular and $42$ range bins. The leftmost approach is the reference oversampling. Next are the interpolation approaches, \gls{dft} and spline, based on the $9$ minimal \gls{dft} samples. The discontinuous periodogram of the strongest peaks extracted by \gls{omp} is the rightmost. The ground truth positions of the two reflectors are indicated by \ref{plot_gt_marker}.}
\label{fig:Intensity_method_overview}
\vspace{-4mm}
\end{figure*}

For the \textit{near} scenario of two octahedral radar reflectors with an inter-object distance of \SI{4.70}{m}, a comparison of the evaluated angular acquisition methods is given in Fig.~\ref{fig:Intensity_method_overview}. All the intensity plots over range and \gls{naf} show the strong component of the rear wall, starting at a range of approx. \SI{22}{m}. In the leftmost Fig.~\ref{fig:oversampl}, the intensity plot of the reference oversampled acquisition is displayed. While this approach is our angular performance reference, for a system with confined angular acquisition capabilities, it is not desired to acquire a multitude of angular samples beyond the minimally necessary, as this procedure would impose heavy constraints on the availability for different tasks, e.g., communications. 
In Fig.~\ref{fig:dftinterp} and Fig.~\ref{fig:splineinterp}, the interpolation methods, \gls{dft} and spline, based on the \gls{dft} samples are shown. It is evident that both approaches exhibit a substantial capability to efficiently replicate the angular response of the targets with reasonable precision. However, it is observed that the angular response of weak reflectors is being extended. The primary distinctions between these two approaches are evident in the presence and variation of background noise, with some variation in the prominence of sidelobes.

Lastly, the results of \gls{omp} approach are shown in Fig.~\ref{fig:omp}.
The \gls{omp} approach results in a different type of periodogram in comparison to the other approaches. \gls{omp} provides the estimated positions of scatterers and their relative power resulting in a sparse, discontinuous range-angle periodogram.

The combination of two types of reflectors, and four unique inter-object spacings results in a total of 8 evaluated scenarios.

\subsection{Ground truth}

For the numerical evaluation, we calculate the \gls{rmse} between the ground truth and the estimates of the corresponding algorithm. 
In the absence of accurate ground truth information, the ground truth positions of the two targets are estimated from the oversampled signal based on separate estimates from the full set of 24 frames recorded for each direction. Specifically, for a single frame in each direction, a single estimate is obtained, and the angular ground truths of the two targets are estimated by taking the median over the individual 24 estimates.

\subsection{Target detection}

The estimation of target positions is initiated with the implementation of a \gls{cfar} algorithm, characterized by a desired probability of false alarm ${P^{\mathrm{FA}}=10^{-6}}$, based on the full range of the periodogram. By focusing on angular estimation, the range domain is constrained to the range bin that exhibits the maximum intensity. In this instance, prior knowledge of the surrounding context is employed by excluding the range of the rear wall.

Finally, we iteratively extract the maximum peaks. Within the system's angular resolution range, only one peak can exist. Therefore, locating a peak limits the potential angular range for subsequent peak detection based on the angular resolution.

\section{Results}

The evaluation of our angular measurement results is presented in two parts. Firstly, the angle-range intensity plots of the different approaches will be examined from a qualitative, visual point of view. We then proceed to a numerical error evaluation with the calculation of the \gls{rmse}.

\subsection{Visual evaluation}

For the different angular estimation methods presented in Fig.~\ref{fig:Intensity_scenario_comparison}, we focus on the range of interest, from \SI{15}{m} to \SI{20}{m}, where our two reflectors are located. The intensity plots over the limited range and full angular sweep range for five different approaches are arranged from top to bottom: oversampling, \gls{dft}-sampled, \gls{dft}-interpolated, spline-interpolated, and \gls{omp}.
The scenarios become more challenging from left to right, i.e., the inter-object spacing increases in the same order. The top row of figures shows the scenarios of the octahedral radar reflectors, which have a significantly smaller reflective area in comparison to the wall reflectors, depicted in the bottom row.

In the \textit{resolution-limited} scenarios, the system as a whole, and thus, all approaches are challenged discriminating the two distinct targets. As the distance between the two reflectors is close to the resolution limits of the system, we observe an interfered angular response resulting in the almost blinding of T2 in Fig~\ref{fig:reflect_314}.
Based on the \gls{dft}-samples, both the \gls{dft} and the spline interpolation achieve good results in reconstructing the oversampled angular response.
The \gls{omp}-based reconstruction provides a sparse and discontinuous angular periodogram, emphasizing the strongest reflections in the scene. 
While achieving high localization accuracy for dominant targets, performance can degrade when scatterers are weak or closely spaced due to the discrete grid assumption and the absence of phase information.
Moreover, the modeling of the dictionary atoms is significantly changing the performance of \gls{omp}.
Nevertheless, in scenarios dominated by a few strong reflections, \gls{omp} serves as an effective low-complexity alternative to interpolation-based reconstruction methods, particularly for analog or hybrid systems with limited observation diversity.

As the inter-object spacing increases, it becomes easier to discriminate the two target peaks in the intensity plots with the naked eye. In general, it seems that the weaker the target, the greater the extent to which the interpolation approaches extend the targets' angular width \wrt the reference oversampled response.
The second increment just operates on the limits of the resolution capabilities of the setup, still causing some interference between the two objects. The last two increments should be increasingly less challenging for the system.

\begin{figure*}[ht]
    \def \showMarker {1}
    \def \showObjects {0}
    \def \selectSuffix {cm_concatenated.png}
    \def \selectFrames {6frames/}
    \def \selectBDir {Data/RangeAngle/}
    \def \selectScenario {angtest_sm_reflectors_}
    \def\subfigwidth{0.224} 
    \def\figwidth{4.9cm}
    \def\figheight{\figwidth}
    \def\genvspace{-1.0mm}
    \def\genhspace{-0.2mm}  
    \centering
    \hspace*{-12.5mm}
    \def \selectDist {314}
    \begin{subfigure}[t]{\subfigwidth\textwidth}  
        \centering
        \pgfplotsset{scaled y ticks=false}

\begin{tikzpicture}

    \begin{axis}[
        axis on top,
        width=\figwidth,
        height=\figheight,
        xmajorgrids=true,
        grid style={solid, black!15},
        enlargelimits=false,
        xmin=-0.25, xmax=0.25,
        xlabel near ticks,
        x label style={yshift=0.2em},
        ymin=0, ymax=50,
        ytick={5,15,25,35,45},
        xtick={-0.3,-0.2,-0.1,0,0.1,0.2,0.3},
        yticklabels={\tiny{\gls{omp}}, \tiny{Spline}, \tiny{\gls{dft}}, \tiny{Samples}, \tiny{Over.}},
        yticklabel style={rotate=90},
        ylabel near ticks,
        xlabel={Azimuth in \gls{naf}},
        y label style={yshift=-0.5em},
        label style={font=\footnotesize},
        tick label style={font=\scriptsize},
        legend cell align={left},
        legend columns = 3,
        legend style={
          fill opacity=1.0,
          draw opacity=1,
          text opacity=1,
          at={(0.999,0.999)},  
          anchor=north east,
          draw=lightgray,
          font=\scriptsize
        },
        ]

        \addplot[forget plot] graphics[
          xmin=-0.2667, 
          xmax=0.2667, 
          ymin=0, 
          ymax=50,
          includegraphics={
            trim= 20 4 20 4,
            clip,}
            ] {\selectBDir\selectFrames\selectScenario\selectDist\selectSuffix};

        
    
        

    \end{axis}

\end{tikzpicture}
        \vspace*{\genvspace}
        \caption{Reflectors with \SI{3.14}{m} spacing.}
        \label{fig:reflect_314}
    \end{subfigure}
    \hspace*{\genhspace}
    \def \selectDist {470}
    \begin{subfigure}[t]{\subfigwidth\textwidth}  
        \centering
        \pgfplotsset{scaled y ticks=false}

\begin{tikzpicture}

    \begin{axis}[
        axis on top,
        width=\figwidth,
        height=\figheight,
        xmajorgrids=true,
        grid style={solid, black!15},
        enlargelimits=false,
        xmin=-0.25, xmax=0.25,
        xlabel near ticks,
        x label style={yshift=0.2em},
        ymin=0, ymax=50,
        ytick={5,15,25,35,45},
        xtick={-0.3,-0.2,-0.1,0,0.1,0.2,0.3},
        yticklabels={\tiny{\gls{omp}}, \tiny{Spline}, \tiny{\gls{dft}}, \tiny{Samples}, \tiny{Over.}},
        yticklabel style={rotate=90},
        ylabel near ticks,
        xlabel={Azimuth in \gls{naf}},
        y label style={yshift=-0.5em},
        label style={font=\footnotesize},
        tick label style={font=\scriptsize},
        legend cell align={left},
        legend columns = 3,
        legend style={
          fill opacity=1.0,
          draw opacity=1,
          text opacity=1,
          at={(0.999,0.999)},  
          anchor=north east,
          draw=lightgray,
          font=\scriptsize
        },
        ]

        \addplot[forget plot] graphics[
          xmin=-0.2667, 
          xmax=0.2667, 
          ymin=0, 
          ymax=50,
          includegraphics={
            trim= 20 4 20 4,
            clip,}
            ] {\selectBDir\selectFrames\selectScenario\selectDist\selectSuffix};

        
    
        

    \end{axis}

\end{tikzpicture}
        \vspace*{\genvspace}
        \caption{Reflectors with \SI{4.70}{m} spacing.}
        \label{fig:reflect_470}
    \end{subfigure}
    \hspace*{\genhspace}
    \def \selectDist {625}
    \begin{subfigure}[t]{\subfigwidth\textwidth}  
        \centering
        \pgfplotsset{scaled y ticks=false}

\begin{tikzpicture}

    \begin{axis}[
        axis on top,
        width=\figwidth,
        height=\figheight,
        xmajorgrids=true,
        grid style={solid, black!15},
        enlargelimits=false,
        xmin=-0.25, xmax=0.25,
        xlabel near ticks,
        x label style={yshift=0.2em},
        ymin=0, ymax=50,
        ytick={5,15,25,35,45},
        xtick={-0.3,-0.2,-0.1,0,0.1,0.2,0.3},
        yticklabels={\tiny{\gls{omp}}, \tiny{Spline}, \tiny{\gls{dft}}, \tiny{Samples}, \tiny{Over.}},
        yticklabel style={rotate=90},
        ylabel near ticks,
        xlabel={Azimuth in \gls{naf}},
        y label style={yshift=-0.5em},
        label style={font=\footnotesize},
        tick label style={font=\scriptsize},
        legend cell align={left},
        legend columns = 3,
        legend style={
          fill opacity=1.0,
          draw opacity=1,
          text opacity=1,
          at={(0.999,0.999)},  
          anchor=north east,
          draw=lightgray,
          font=\scriptsize
        },
        ]

        \addplot[forget plot] graphics[
          xmin=-0.2667, 
          xmax=0.2667, 
          ymin=0, 
          ymax=50,
          includegraphics={
            trim= 20 4 20 4,
            clip,}
            ] {\selectBDir\selectFrames\selectScenario\selectDist\selectSuffix};

        
    
        

    \end{axis}

\end{tikzpicture}
        \vspace*{\genvspace}
        \caption{Reflectors with \SI{6.25}{m} spacing.}
        \label{fig:reflect_625}
    \end{subfigure}
    \hspace*{\genhspace}
    \def \selectDist {779}
    \begin{subfigure}[t]{\subfigwidth\textwidth}  
        \centering
        \pgfplotsset{scaled y ticks=false}

\begin{tikzpicture}

    \begin{axis}[
        axis on top,
        width=\figwidth,
        height=\figheight,
        xmajorgrids=true,
        grid style={solid, black!15},
        enlargelimits=false,
        xmin=-0.25, xmax=0.25,
        xlabel near ticks,
        x label style={yshift=0.2em},
        ymin=0, ymax=50,
        ytick={5,15,25,35,45},
        xtick={-0.3,-0.2,-0.1,0,0.1,0.2,0.3},
        yticklabels={\tiny{\gls{omp}}, \tiny{Spline}, \tiny{\gls{dft}}, \tiny{Samples}, \tiny{Over.}},
        yticklabel style={rotate=90},
        ylabel near ticks,
        xlabel={Azimuth in \gls{naf}},
        y label style={yshift=-0.5em},
        label style={font=\footnotesize},
        tick label style={font=\footnotesize},
        legend cell align={left},
        legend columns = 3,
        legend style={
          fill opacity=1.0,
          draw opacity=1,
          text opacity=1,
          at={(0.999,0.999)},  
          anchor=north east,
          draw=lightgray,
          font=\scriptsize
        },
        colorbar,
        point meta min=-60,
        point meta max=0,
        colormap name=jet_inue,
        colorbar style=
            {ylabel={Power in dB}, 
            at={(1.08, 0)},
            anchor = south,
            width = 0.2cm,
            ytick={-60,-45,-15, 0},
            ylabel style={yshift=0.8cm},
            every axis/.append style=
                {font=\footnotesize}
            }
        ]

        \addplot[forget plot] graphics[
          xmin=-0.2667, 
          xmax=0.2667, 
          ymin=0, 
          ymax=50,
          includegraphics={
            trim= 20 4 20 4,
            clip,}
            ] {\selectBDir\selectFrames\selectScenario\selectDist\selectSuffix};

        
    
        

    \end{axis}

\end{tikzpicture}
        \vspace*{\genvspace}
        \vspace{-4.2mm}
        \caption{Reflectors with \SI{7.79}{m} spacing.}
        \label{fig:reflect_779}
    \end{subfigure}

    \hspace*{-12.5mm}
    \def \selectScenario {angtest_walls_}
    \def \selectDist {314}
    \begin{subfigure}[t]{\subfigwidth\textwidth}  
        \centering
        \pgfplotsset{scaled y ticks=false}

\begin{tikzpicture}

    \begin{axis}[
        axis on top,
        width=\figwidth,
        height=\figheight,
        xmajorgrids=true,
        grid style={solid, black!15},
        enlargelimits=false,
        xmin=-0.25, xmax=0.25,
        xlabel near ticks,
        x label style={yshift=0.2em},
        ymin=0, ymax=50,
        ytick={5,15,25,35,45},
        xtick={-0.3,-0.2,-0.1,0,0.1,0.2,0.3},
        yticklabels={\tiny{\gls{omp}}, \tiny{Spline}, \tiny{\gls{dft}}, \tiny{Samples}, \tiny{Over.}},
        yticklabel style={rotate=90},
        ylabel near ticks,
        xlabel={Azimuth in \gls{naf}},
        y label style={yshift=-0.5em},
        label style={font=\footnotesize},
        tick label style={font=\scriptsize},
        legend cell align={left},
        legend columns = 3,
        legend style={
          fill opacity=1.0,
          draw opacity=1,
          text opacity=1,
          at={(0.999,0.999)},  
          anchor=north east,
          draw=lightgray,
          font=\scriptsize
        },
        ]

        \addplot[forget plot] graphics[
          xmin=-0.2667, 
          xmax=0.2667, 
          ymin=0, 
          ymax=50,
          includegraphics={
            trim= 20 4 20 4,
            clip,}
            ] {\selectBDir\selectFrames\selectScenario\selectDist\selectSuffix};

        
    
        

    \end{axis}

\end{tikzpicture}
        \vspace*{\genvspace}
        \caption{Walls with \SI{3.14}{m} spacing.}
        \label{fig:walls_314}
    \end{subfigure}
    \hspace*{\genhspace}
    \def \selectDist {470}
    \begin{subfigure}[t]{\subfigwidth\textwidth}  
        \centering
        \pgfplotsset{scaled y ticks=false}

\begin{tikzpicture}

    \begin{axis}[
        axis on top,
        width=\figwidth,
        height=\figheight,
        xmajorgrids=true,
        grid style={solid, black!15},
        enlargelimits=false,
        xmin=-0.25, xmax=0.25,
        xlabel near ticks,
        x label style={yshift=0.2em},
        ymin=0, ymax=50,
        ytick={5,15,25,35,45},
        xtick={-0.3,-0.2,-0.1,0,0.1,0.2,0.3},
        yticklabels={\tiny{\gls{omp}}, \tiny{Spline}, \tiny{\gls{dft}}, \tiny{Samples}, \tiny{Over.}},
        yticklabel style={rotate=90},
        ylabel near ticks,
        xlabel={Azimuth in \gls{naf}},
        y label style={yshift=-0.5em},
        label style={font=\footnotesize},
        tick label style={font=\scriptsize},
        legend cell align={left},
        legend columns = 3,
        legend style={
          fill opacity=1.0,
          draw opacity=1,
          text opacity=1,
          at={(0.999,0.999)},  
          anchor=north east,
          draw=lightgray,
          font=\scriptsize
        },
        ]

        \addplot[forget plot] graphics[
          xmin=-0.2667, 
          xmax=0.2667, 
          ymin=0, 
          ymax=50,
          includegraphics={
            trim= 20 4 20 4,
            clip,}
            ] {\selectBDir\selectFrames\selectScenario\selectDist\selectSuffix};

        
    
        

    \end{axis}

\end{tikzpicture}
        \vspace*{\genvspace}
        \caption{Walls with \SI{4.70}{m} spacing.}
        \label{fig:walls_470}
    \end{subfigure}
    \hspace*{\genhspace}
    \def \selectDist {625}
    \begin{subfigure}[t]{\subfigwidth\textwidth}  
        \centering
        \pgfplotsset{scaled y ticks=false}

\begin{tikzpicture}

    \begin{axis}[
        axis on top,
        width=\figwidth,
        height=\figheight,
        xmajorgrids=true,
        grid style={solid, black!15},
        enlargelimits=false,
        xmin=-0.25, xmax=0.25,
        xlabel near ticks,
        x label style={yshift=0.2em},
        ymin=0, ymax=50,
        ytick={5,15,25,35,45},
        xtick={-0.3,-0.2,-0.1,0,0.1,0.2,0.3},
        yticklabels={\tiny{\gls{omp}}, \tiny{Spline}, \tiny{\gls{dft}}, \tiny{Samples}, \tiny{Over.}},
        yticklabel style={rotate=90},
        ylabel near ticks,
        xlabel={Azimuth in \gls{naf}},
        y label style={yshift=-0.5em},
        label style={font=\footnotesize},
        tick label style={font=\scriptsize},
        legend cell align={left},
        legend columns = 3,
        legend style={
          fill opacity=1.0,
          draw opacity=1,
          text opacity=1,
          at={(0.999,0.999)},  
          anchor=north east,
          draw=lightgray,
          font=\scriptsize
        },
        ]

        \addplot[forget plot] graphics[
          xmin=-0.2667, 
          xmax=0.2667, 
          ymin=0, 
          ymax=50,
          includegraphics={
            trim= 20 4 20 4,
            clip,}
            ] {\selectBDir\selectFrames\selectScenario\selectDist\selectSuffix};

        
    
        

    \end{axis}

\end{tikzpicture}
        \vspace*{\genvspace}
        \caption{Walls with \SI{6.25}{m} spacing.}
        \label{fig:walls_635}
    \end{subfigure}
    \hspace*{\genhspace}
    \def \selectDist {779}
    \begin{subfigure}[t]{\subfigwidth\textwidth}  
        \centering
        \pgfplotsset{scaled y ticks=false}

\begin{tikzpicture}

    \begin{axis}[
        axis on top,
        width=\figwidth,
        height=\figheight,
        xmajorgrids=true,
        grid style={solid, black!15},
        enlargelimits=false,
        xmin=-0.25, xmax=0.25,
        xlabel near ticks,
        x label style={yshift=0.2em},
        ymin=0, ymax=50,
        ytick={5,15,25,35,45},
        xtick={-0.3,-0.2,-0.1,0,0.1,0.2,0.3},
        yticklabels={\tiny{\gls{omp}}, \tiny{Spline}, \tiny{\gls{dft}}, \tiny{Samples}, \tiny{Over.}},
        yticklabel style={rotate=90},
        ylabel near ticks,
        xlabel={Azimuth in \gls{naf}},
        y label style={yshift=-0.5em},
        label style={font=\footnotesize},
        tick label style={font=\footnotesize},
        legend cell align={left},
        legend columns = 3,
        legend style={
          fill opacity=1.0,
          draw opacity=1,
          text opacity=1,
          at={(0.999,0.999)},  
          anchor=north east,
          draw=lightgray,
          font=\scriptsize
        },
        colorbar,
        point meta min=-60,
        point meta max=0,
        colormap name=jet_inue,
        colorbar style=
            {ylabel={Power in dB}, 
            at={(1.08, 0)},
            anchor = south,
            width = 0.2cm,
            ytick={-60,-45,-15, 0},
            ylabel style={yshift=0.8cm},
            every axis/.append style=
                {font=\footnotesize}
            }
        ]

        \addplot[forget plot] graphics[
          xmin=-0.2667, 
          xmax=0.2667, 
          ymin=0, 
          ymax=50,
          includegraphics={
            trim= 20 4 20 4,
            clip,}
            ] {\selectBDir\selectFrames\selectScenario\selectDist\selectSuffix};

        
    
        

    \end{axis}

\end{tikzpicture}
        \vspace*{\genvspace}
        \vspace{-4.2mm}
        \caption{Walls with \SI{7.79}{m} spacing.}
        \label{fig:walls_779}
    \end{subfigure}
    \caption {Angular acquisition or reconstruction comparison over full set of scenarios. The inter-object spacing of the scenarios is increasing from left to right (\textit{resolution-limited}: \SI{3.14}{m} $\sim$ \SI{10}{^{\circ}},  \textit{near}: \SI{4.70}{m} $\sim$ \SI{15}{^{\circ}},  \textit{mid}: \SI{6.25}{m} $\sim$ \SI{20}{^{\circ}}, \textit{far}: \SI{7.79}{m} $\sim$ \SI{25}{^{\circ}}). In the top row, a) to d), are the scenarios of the two octahedral reflectors, whereas in the bottom row, e) to h), are the scenarios of the two wall reflectors. The range of each figure is constrained to the interval of interest, from \SI{15}{m} to \SI{20}{m}, where the two targets are located. In this range, the angular responses of the various approaches are arranged in a vertical stack.}

\label{fig:Intensity_scenario_comparison}
\vspace{-4mm}
\end{figure*}
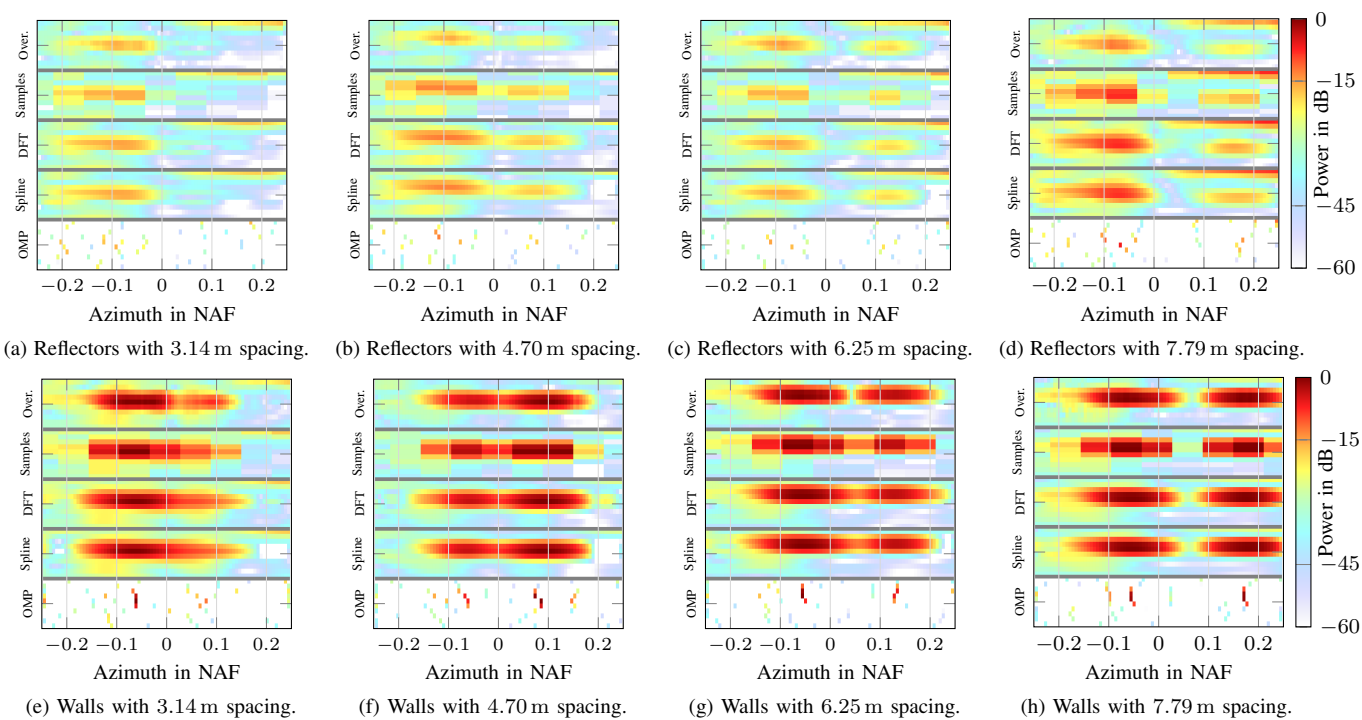

\subsection{Numerical evaluation}

\def \selectErrorDir {Data/ErrorEval/6frames/}
\def \selectCSV {angtest_sm_reflectors_rmse.txt}
\pgfplotstableread[col sep=comma]{\selectErrorDir\selectCSV}\datatableFirst
        
\def \selectRowIdx {3}
\pgfplotstablegetelem{\selectRowIdx}{[index]0}\of\datatableFirst
\let\errorValFirstTargetDFTFirstRow\pgfplotsretval

\pgfplotstablegetelem{\selectRowIdx}{[index]1}\of\datatableFirst
\let\errorValSecondTargetDFTFirstRow\pgfplotsretval

\pgfplotstablegetelem{\selectRowIdx}{[index]2}\of\datatableFirst
\let\errorValCombinedDFTFirstRow\pgfplotsretval

\def \selectRowIdx {1}
\pgfplotstablegetelem{\selectRowIdx}{[index]0}\of\datatableFirst
\let\errorValFirstTargetSplineFirstRow\pgfplotsretval

\pgfplotstablegetelem{\selectRowIdx}{[index]1}\of\datatableFirst
\let\errorValSecondTargetSplineFirstRow\pgfplotsretval

\pgfplotstablegetelem{\selectRowIdx}{[index]2}\of\datatableFirst
\let\errorValCombinedSplineFirstRow\pgfplotsretval

\def \selectRowIdx {2}
\pgfplotstablegetelem{\selectRowIdx}{[index]0}\of\datatableFirst
\let\errorValFirstTargetOMPFirstRow\pgfplotsretval

\pgfplotstablegetelem{\selectRowIdx}{[index]1}\of\datatableFirst
\let\errorValSecondTargetOMPFirstRow\pgfplotsretval

\pgfplotstablegetelem{\selectRowIdx}{[index]2}\of\datatableFirst
\let\errorValCombinedOMPFirstRow\pgfplotsretval

\def \tabFirstRow {\textcolor{blue}{\errorValFirstTargetDFTFirstRow} \textcolor{teal}{\errorValFirstTargetSplineFirstRow} \textcolor{violet}{\errorValFirstTargetOMPFirstRow} & \textcolor{blue}{\errorValSecondTargetDFTFirstRow} \textcolor{teal}{\errorValSecondTargetSplineFirstRow} \textcolor{violet}{\errorValSecondTargetOMPFirstRow}}

\def \selectCSV {angtest_walls_rmse.txt}
\pgfplotstableread[col sep=comma]{\selectErrorDir\selectCSV}\datatableSecond
        
\def \selectRowIdx {3}
\pgfplotstablegetelem{\selectRowIdx}{[index]0}\of\datatableSecond
\let\errorValFirstTargetDFTThirdRow\pgfplotsretval

\pgfplotstablegetelem{\selectRowIdx}{[index]1}\of\datatableSecond
\let\errorValSecondTargetDFTThirdRow\pgfplotsretval

\pgfplotstablegetelem{\selectRowIdx}{[index]2}\of\datatableSecond
\let\errorValCombinedDFTThirdRow\pgfplotsretval

\def \selectRowIdx {1}
\pgfplotstablegetelem{\selectRowIdx}{[index]0}\of\datatableSecond
\let\errorValFirstTargetSplineThirdRow\pgfplotsretval

\pgfplotstablegetelem{\selectRowIdx}{[index]1}\of\datatableSecond
\let\errorValSecondTargetSplineThirdRow\pgfplotsretval

\pgfplotstablegetelem{\selectRowIdx}{[index]2}\of\datatableSecond
\let\errorValCombinedSplineThirdRow\pgfplotsretval

\def \selectRowIdx {2}
\pgfplotstablegetelem{\selectRowIdx}{[index]0}\of\datatableSecond
\let\errorValFirstTargetOMPThirdRow\pgfplotsretval

\pgfplotstablegetelem{\selectRowIdx}{[index]1}\of\datatableSecond
\let\errorValSecondTargetOMPThirdRow\pgfplotsretval

\pgfplotstablegetelem{\selectRowIdx}{[index]2}\of\datatableSecond
\let\errorValCombinedOMPThirdRow\pgfplotsretval

\def \tabThirdRow {\textcolor{blue}{\errorValFirstTargetDFTThirdRow} \textcolor{teal}{\errorValFirstTargetSplineThirdRow} \textcolor{violet}{\errorValFirstTargetOMPThirdRow} & \textcolor{blue}{\errorValSecondTargetDFTThirdRow} \textcolor{teal}{\errorValSecondTargetSplineThirdRow} \textcolor{violet}{\errorValSecondTargetOMPThirdRow}}

\def \selectCSV {angtest_combined_rmse.txt}
\pgfplotstableread[col sep=comma]{\selectErrorDir\selectCSV}\datatableFifth
        
\def \selectRowIdx {3}
\pgfplotstablegetelem{\selectRowIdx}{[index]0}\of\datatableFifth
\let\errorValFirstTargetDFTFifthRow\pgfplotsretval

\pgfplotstablegetelem{\selectRowIdx}{[index]1}\of\datatableFifth
\let\errorValSecondTargetDFTFifthRow\pgfplotsretval

\pgfplotstablegetelem{\selectRowIdx}{[index]2}\of\datatableFifth
\let\errorValCombinedDFTFifthRow\pgfplotsretval

\def \selectRowIdx {1}
\pgfplotstablegetelem{\selectRowIdx}{[index]0}\of\datatableFifth
\let\errorValFirstTargetSplineFifthRow\pgfplotsretval

\pgfplotstablegetelem{\selectRowIdx}{[index]1}\of\datatableFifth
\let\errorValSecondTargetSplineFifthRow\pgfplotsretval

\pgfplotstablegetelem{\selectRowIdx}{[index]2}\of\datatableFifth
\let\errorValCombinedSplineFifthRow\pgfplotsretval

\def \selectRowIdx {2}
\pgfplotstablegetelem{\selectRowIdx}{[index]0}\of\datatableFifth
\let\errorValFirstTargetOMPFifthRow\pgfplotsretval

\pgfplotstablegetelem{\selectRowIdx}{[index]1}\of\datatableFifth
\let\errorValSecondTargetOMPFifthRow\pgfplotsretval

\pgfplotstablegetelem{\selectRowIdx}{[index]2}\of\datatableFifth
\let\errorValCombinedOMPFifthRow\pgfplotsretval

\def \tabFifthRow {\textcolor{blue}{\errorValFirstTargetDFTFifthRow} \textcolor{teal}{\errorValFirstTargetSplineFifthRow} \textcolor{violet}{\errorValFirstTargetOMPFifthRow} & \textcolor{blue}{\errorValSecondTargetDFTFifthRow} \textcolor{teal}{\errorValSecondTargetSplineFifthRow} \textcolor{violet}{\errorValSecondTargetOMPFifthRow}}

\def \tabLastRow {\textbf{\textcolor{blue}{\errorValCombinedDFTFifthRow} \textcolor{teal}{\errorValCombinedSplineFifthRow}  \textcolor{violet}{\errorValCombinedOMPFifthRow}}}

\renewcommand{\arraystretch}{1.30} 
\begin{table}
    \centering
    \caption{\gls{rmse} in \gls{naf}}
    \label{tab:poc_rmse}
     \begin{tabular}{|c||c|c|}
        \hline
         & Target \textbf{T1} & Target \textbf{T2} \\
        \Xhline{3\arrayrulewidth}
        Reflectors & \tabFirstRow \\
        \hline
        Walls& \tabThirdRow \\
        \hline
        Combined & \tabFifthRow \\
        \Xhline{3\arrayrulewidth}
        \textbf{Total} & \multicolumn{2}{c|}{\tabLastRow} \\
        \hline
        \hline
         Legend & \multicolumn{2}{c|}{\textcolor{blue}{\gls{dft}} \textcolor{teal}{Spline} \textcolor{violet}{\gls{omp}}} \\
         \hline
    \end{tabular}
\vspace{-4mm}
\end{table}

As a metric for the precision of the angular estimation, we calculate the \gls{naf} \gls{rmse} between the ground truths of the targets and their nearest estimate from the different approaches. Taking the mean of 6 frames, we present the results in Tab.~\ref{tab:poc_rmse} for each target and target type. Additionally, we calculated the variance of errors, which over all methods and scenarios is very stable in the range of $10^{-4}$. While the numerical difference between multiple \gls{rmse} values might appear insignificant within the range of $10^-3$ \gls{naf}, the difference in positional error can still be substantial. For instance, at a target distance of \SI{18}{m} the difference in cross track error between the best and worst approach is approx. \SI{0.3}{m}.

The \gls{rmse} evaluation of the peak estimates shows robust results for the interpolation approaches of \gls{dft} and spline in both reflector type scenarios. \gls{dft}-based interpolation shows overall the best performance, while having also the most stable \gls{rmse} values over any scenario. Spline interpolation achieves close to best performance while having a slightly favorable estimation of the right target T2. 
\gls{omp} is more challenged by the weaker octahedral reflectors, but for the stronger wall reflectors can even provide the best over all estimate for T2.

Generally, for the smaller octahedral reflectors, the left target T1 is detectable within all scenarios, while the right target T2 has more fluctuation. This finding aligns with the observations for the \gls{dft} and \gls{omp} approach.
However, for the bigger wall reflectors, the interference of the two target positions seems to have a greater influence on T1. 

In summary, the evaluation should be considered with a certain degree of caution, given the relatively limited scope of the measured data. It is noteworthy that under optimal conditions, such as sufficient inter-object distance and the presence of reflective properties on the scatterer, \gls{omp} can achieve the most precise single object estimation.
However, the overall tendency that can be extrapolated is that consistent angular estimation, generalizing over different types of targets and scenarios, is easier to achieve with interpolation approaches, with \gls{dft} demonstrating a slight advantage in estimation errors. 

\section{Conclusion}

The present study effectively demonstrates the application of advanced angular estimation methodologies in industrial environments utilizing a 5G NR \gls{poc} system. The underlying processing of OFDM data within the 5G NR framework demonstrates the feasibility of \gls{isac} cellular networks to facilitate precise angular estimation. 

By leveraging a minimal set of angular samples based on the \gls{dft} theory of~\cite{mandelli2022sampling}, the proposed approaches enable systems of confined angular capabilities and optimize computational efficiency.
The evaluation indicates in measurement scenarios characterized by clearly spaced and strong scatterers, the \gls{omp} approach achieves the best estimate for a single target. While the \gls{omp} performance is more dependent on parameterization, this could imply further potential to be unlocked with the other approaches.
Overall, both interpolation approaches, \gls{dft}-based and spline, demonstrate strong proficiency in the reconstruction  of the angular domain up to the system's resolution limits. Specifically, the \gls{dft}-based interpolation demonstrates a high degree of generalization across all scenarios, exhibiting a slight advantage in terms of error performance over spline, and a notable advantage over \gls{omp}.

In subsequent studies, utilizing joint and optimized beamforming coefficients is intended to bridge the gap to the theoretically achievable angular resolution of the physical setup.
Furthermore, we aim to expand the scope of measurements to other environments and different \gls{ru} topologies.

\appendices

\section*{Acknowledgment}
The authors thank Artjom~Grudnitsky and Paolo~Tosi for their support during the development of this work.

This work was developed within the KOMSENS-6G project, partly funded by the German Ministry of Education and Research under grant 16KISK112K and 16KISK120.

\balance

\bibliographystyle{IEEEtran}
\bibliography{references.bib}

\end{NoHyper}
\end{document}